\documentclass[12pt,a4paper]{article}
\input epsf
\usepackage{amsmath,amsfonts,epsfig,color,latexsym}
\usepackage{amssymb}
\usepackage[english, USenglish]{babel}
\usepackage{epsfig}
\usepackage{a4wide}
\usepackage{rotating}
\usepackage{hyperref}
\usepackage{float}

\catcode`\@=11
\newcount\hour
\newcount\minute
\newtoks\amorpm
\hour=\time\divide\hour by60
\minute=\time{\multiply\hour by60 \global\advance\minute by-\hour}
\edef\standardtime{{\ifnum\hour<12 \global\amorpm={am}%
        \else\global\amorpm={pm}\advance\hour by-12 \fi
        \ifnum\hour=0 \hour=12 \fi
        \number\hour:\ifnum\minute<10 0\fi\number\minute\the\amorpm}}
\edef\militarytime{\number\hour:\ifnum\minute<10 0\fi\number\minute}
\def\draftlabel#1{{\@bsphack\if@filesw {\let\thepage\relax
   \xdef\@gtempa{\write\@auxout{\string
      \newlabel{#1}{{\@currentlabel}{\thepage}}}}}\@gtempa
   \if@nobreak \ifvmode\nobreak\fi\fi\fi\@esphack}
        \gdef\@eqnlabel{#1}}
\def\@eqnlabel{}
\def\@vacuum{}
\def\draftmarginnote#1{\marginpar{\raggedright\scriptsize\tt#1}}
\def\draft{\oddsidemargin -.2truein
        \def\@oddfoot{\sl Preliminary Notes \hfil
        \rm\thepage\hfil\sl\today\quad\militarytime}
        \let\@evenfoot\@oddfoot \overfullrule 3pt
        \let\label=\draftlabel
        \let\marginnote=\draftmarginnote
   \def\@eqnnum{(\theequation)\rlap{\kern\marginparsep\tt\@eqnlabel}%
\global\let\@eqnlabel\@vacuum}  }
\relax

\newcommand{\be}{\begin{equation}}
\newcommand{\ee}{\end{equation}}
\newcommand{\bea}{\begin{eqnarray}}
\newcommand{\eea}{\end{eqnarray}}

\newcommand{\cN}{{\cal N}}

\newcommand{\bt}[1]{{\bar t}}

\newcommand{\gr}[1]{\Gamma_{[#1]}}

\newcommand{\f}{\frac}

\newcommand{\bor}{\mathcal{G}}

\newcommand{\we}{\vec{w}_1}
\newcommand{\wz}{\vec{w}_2}
\newcommand{\wv}{\vec{w}}

\newcommand{\y}{{\bf y}}
\newcommand{\grg}{\mathfrak{G}}

\author{\\[-0.3cm] Stefan Hohenegger\footnote{{\tt
shoheneg@mppmu.mpg.de}}~~and  Daniel~Persson\footnote{{\tt
daniel.persson@itp.phys.ethz.ch}}\,\,\,\footnote{Also at {\it Fundamental Physics, Chalmers University of Technology, 412 96 Gothenburg, Sweden.}}}
\title{\begin{flushright}{\vspace{-0.8cm}\small MPP-2011-83}\end{flushright}
\vspace{0.8cm} Enhanced Gauge Groups in $\cN=4$ Topological Amplitudes and Lorentzian Borcherds Algebras}
\dateUSenglish
\date{}
\graphicspath{{figures/}}

\begin{document}

\begin{titlepage}

\maketitle
\thispagestyle{empty}
\begin{center}
\renewcommand{\thefootnote}{\fnsymbol{footnote}}\vspace{-0.5cm}
\footnotemark[1]\emph{Max--Planck--Institut f\"ur
Physik,\\Werner--Heisenberg--Institut,\\
F\"ohringer Ring 6, 80805 M\"unchen, Germany}\\[.4cm]
\vspace{0.1cm}
\footnotemark[2] \emph{Institut f\"ur Theoretische
Physik, ETH
  Z\"urich, \\
CH-8093 Z\"urich, Switzerland}\\[1cm]

\end{center}
\begin{abstract}
\noindent We continue our study of algebraic properties of $\mathcal{N}=4$ topological amplitudes in heterotic string theory compactified on $\mathbb{T}^2$, initiated in arXiv:1102.1821. In this work we evaluate a particular one-loop amplitude for any enhanced gauge group $\mathfrak{h}\subset \mathfrak{e}_8\oplus \mathfrak{e}_8$, i.e. for arbitrary choice of Wilson line moduli. We show that a certain analytic part of the result has an infinite product representation, where the product is taken over the positive roots of a Lorentzian Kac-Moody algebra $\mathfrak{g}^{++}$. The latter is obtained through double extension of the complement $\mathfrak{g}= (\mathfrak{e}_8\oplus \mathfrak{e}_8)/\mathfrak{h}$. The infinite product is automorphic with respect to a finite index subgroup of the full T-duality group $SO(2,18;\mathbb{Z})$ and, through the philosophy of Borcherds-Gritsenko-Nikulin, this defines the denominator formula of a generalized Kac-Moody algebra $\mathcal{G}(\mathfrak{g}^{++})$, which is an 'automorphic correction' of $\mathfrak{g}^{++}$. We explicitly give the root multiplicities of $\mathcal{G}(\mathfrak{g}^{++})$ for a number of examples.

\end{abstract}
\end{titlepage}
\noindent

\setcounter{page}{2}
\newpage 
\tableofcontents
\newpage 
\renewcommand{\theequation}{\arabic{section}.\arabic{equation}}
\section{Introduction}

This work is a continuation of our previous analysis \cite{GHPI} of a particular class of BPS-saturated higher string-loop amplitudes $\mathcal{F}_g$, arising in type II string theory compactified  on K3$\times {\mathbb T}^2$, which are captured by correlation functions of the $\mathcal{N}=4$ topological string \cite{Antoniadis:2006mr}. For any loop order $g$, the dual amplitudes in heterotic string theory compactified on ${\mathbb T}^6$ receive contributions at all loop orders in (heterotic) perturbation theory. However, the leading contribution in the weak coupling limit is a one-loop expression which can be studied in detail. In particular, following a similar analysis of Harvey and Moore in the $\mathcal{N}=2$ setting \cite{Harvey:1995fq,Harvey:1996gc}, one can explicitly perform the worldsheet  torus integral and analyse its algebraic and modular properties. The latter are expected as a consequence of the familiar worldsheet  $SL(2,\mathbb{Z})$-invariance of the integrand. Mathematically, heterotic one-loop amplitudes fall into the category of so called `singular theta correspondences', as analysed in detail by Borcherds \cite{Borcherds1,Borcherds2}. This means that the worldsheet  $SL(2,\mathbb{Z})$ and the target space T-duality group $SO(6,22;\mathbb{Z})$ form a dual reductive pair within a larger metaplectic group, and integrating over the fundamental domain of $SL(2,\mathbb{Z})$ thus induces good modular properties under the T-duality group.

We focus on an analytic piece of the one-loop integral which is characterised by the fact that it does not violate a particular class of supersymmetric Ward identities discussed in \cite{Antoniadis:2006mr,Antoniadis:2007cw} (see also \cite{Berkovits:1994vy,Ooguri:1991fp}). Upon splitting $\mathbb{T}^6=\mathbb{T}^4\times \mathbb{T}^2$, we evaluate $\mathcal{F}_{1}^{\text{analy}}$ explicitly in the large volume limit of $\mathbb{T}^4$ using the method of orbits \cite{Dixon:1990pc,Harvey:1995fq} (or in mathematical parlance, the Rankin-Selberg method), for any choice of the unbroken gauge group $\mathfrak{h}\subset \mathfrak{e}_8\oplus \mathfrak{e}_8$. In particular, we allow for $\mathfrak{h}$ to be semisimple, thus extending previous results in the literature where always one $\mathfrak{e}_8$-factor remained unbroken \cite{Harvey:1995fq,Henningson:1996jz,Marino:1998pg,Weiss:2007tk, LopesCardoso:1996nc,Stieberger:1998yi}. Equivalently, we allow for Wilson lines to be embedded in any of the factors in the sum $\mathfrak{e}_8\oplus \mathfrak{e}_8$. As we will see, in this setting the theta correspondence is generalised to subgroups of $SL(2,\mathbb{Z})$ and the T-duality group respectively.

Taking the above mentioned modular properties into account, we show further that part of the analytic integral $\mathcal{F}_{1}^{\text{analy}}$ can be written as an infinite product over the positive root lattice of the Lorentzian Kac-Moody algebra $\mathfrak{g}^{++}$, which is a double extension of the complement $\mathfrak{g}=(\mathfrak{e}_8\oplus \mathfrak{e}_8)/\mathfrak{h}$. In the spirit of  \cite{Gritsenko:1996ax,GritsenkoNikulin}, the infinite product so obtained then defines an `automorphic correction' $\mathcal{G}(\mathfrak{g}^{++})$ of $\mathfrak{g}^{++}$. The correction $\mathcal{G}(\mathfrak{g}^{++})$ is a Borcherds-Kac-Moody (BKM) algebra with real root lattice coinciding with the root lattice of $\mathfrak{g}^{++}$. The BPS-integral $\mathcal{F}_1^{\text{analy}}$ can then be related to the denominator formula of the BKM-algebra $\mathcal{G}(\mathfrak{g}^{++})$.

This paper is structured as follows. In section \ref{NarainComp} we discuss some relevant features of Wilson line moduli in Narain compactifications, with special emphasis on how the splitting of the Narain lattice $\Gamma^{6,22}$ depends on the enhanced gauge group $\mathfrak{h}\subset \mathfrak{e}_8\oplus \mathfrak{e}_8$. Then, in section \ref{Section:Akintegral}, we first recall the structure of the $\mathcal{N}=4$ topological amplitudes $\mathcal{F}_g$, focusing on the role of harmonicity, which allows us to  single out the analytic part $\mathcal{F}_g^{\text{analy}}$ of the full amplitude $\mathcal{F}_g$. In section \ref{Sect:TorusIntegralEval} we evaluate the one-loop integral $\mathcal{F}_{1}^{\text{analy}}$ explicitly, and in section \ref{Sect:DenominatorFormula} we show how to write part of the result in terms of an infinite product, from which we extract the denominator formula $\Phi_\mathfrak{g}(\y)$ of the BKM-algebra $\mathcal{G}(\mathfrak{g}^{++})$. We show that the automorphic properties of the denominator formula with respect to a subgroup $\grg \subset SO(2, 18;\mathbb{Z})$ can be made manifest through an integral representation of $\Phi_{\mathfrak{g}}(\y)$, corresponding to a theta correspondence for certain congruence subgroups $\grg\times \Gamma_{[0]} \subset SO(2,18;\mathbb{Z})\times SL(2,\mathbb{Z})$. In section~\ref{Sect:Examples} we further analyse some particular examples in detail. Finally, we end in section~\ref{Conclusions} with a discussion of our results and suggestions for future work. Various calculational details and some relevant mathematical background are relegated to the three Appendices \ref{App:BorcherdsExtension}-\ref{App:ModFormsG04}.
%%%%%%%%%%%%%%%%%%%%%%%%%%%%%%%%%%%%%%%%%%%%%%%%%%%%%%%%%%%%%%%%%%%%%%%%%%%%%%%%%%%%%%%%%%%%%%%%%%%%%%%%%
%%%%%%%%%%%%%%%%%%%%%%%%%%%%%%%%%%%%%%%%%%%%%%%%%%%%%%%%%%%%%%%%%%%%%%%%%%%%%%%%%%%%%%%%%%%%%%%%%%%%%%%%%
\section{Wilson Line Moduli in  Narain Compactifications}
\label{NarainComp}
We start by reviewing some basic facts about toroidal compactifications of the $E_8\times E_8$ heterotic string. 
The classical moduli space for heterotic string theory on ${\mathbb T}^6$ is
described by the coset space
\be\label{classmod}
\mathcal{M}= \Bigl( SL(2,\mathbb{R})/U(1)  \Bigr) \, \times \, \Bigl(
SO(6,22)/(SO(6)\times SO(22))\Bigr)\ ,
\ee
where the first factor encodes the heterotic `axio-dilaton', while the second
factor 
accounts for the remaining Narain moduli of the torus. In order to make contact with the topological amplitudes analysed in section~\ref{Section:Akintegral}, we only need to consider the perturbative string spectrum. The latter consists of the states that are created from a momentum ground state labeled by 
$(p^L,\vec{p};p^R,\vec{p})$ by the action of the oscillators. Here the
compactified (and internal) 
momenta  take values in the Narain lattice\footnote{In our conventions, the left-movers are `supersymmetric', while the right-movers are `bosonic'.}, $(p^{L}, p^{R}) \in \Gamma^{6,22}$, while $\vec{p}$ describes the space-time momentum, {\it i.e.} the uncompactified 4-dimensional
theory. The T-duality group which leaves the Narain lattice invariant is $SO(6,22;\mathbb{Z})$, and thus the
quantum moduli space is the quotient of (\ref{classmod}) by this arithmetic
group. 

For the one-loop string amplitudes it will be important to realize that the BPS states come in two different classes:  Since the compactification of 
heterotic string theory on ${\mathbb T}^6$ preserves 
$\mathcal{N}=4$ supersymmetry we may distinguish between $1/2$ BPS-states associated with short multiplets, and
$1/4$ BPS-states associated 
with intermediate multiplets. As has been discussed in \cite{Lerche:1999ju} all $1/4$ BPS-states are non-perturbative and only the $1/2$ BPS-states are perturbative. Thus, perturbative topological amplitudes only receive contributions from $1/2$ BPS states.

%%%%%%%%%%%%%%%%%%%%%%%%%%%%%%%%%%%%%%%%%%%%%%%%%%%%%%%%%%%%%%%%%%%%%%%%%%%%%%%%%%%%%%%%%
\subsection{The Sublattice $\Gamma_{{\mathfrak g}}\subset \Gamma^{6,22}$}\label{sec:sublattice}

In the following we shall be interested in a certain sublattice $\Gamma_{{\mathfrak g}}$ of the full momentum
lattice
$\Gamma^{6,22}$. In order to describe this
sublattice, we proceed
in two steps. First we split ${\mathbb T}^6={\mathbb T}^2\times {\mathbb T}^4$, 
and take the large-volume limit of the ${\mathbb T}^4$,  effectively setting
the 
${\mathbb T}^4$ momenta to zero. This corresponds to restricting
ourselves to momentum ground states in the even self-dual lattice
$\Gamma^{2,18}$ of signature 
$(2,18)$, which is obtained by splitting
\begin{equation}
\Gamma^{6,22} = \Gamma^{2,18} \oplus \Gamma^{4,4} \ ,\label{LatticeSplitLargeVolume}
\end{equation}
where $\Gamma^{4,4}$ describes the momenta of the ${\mathbb T}^4$. Notice that we are therefore effectively considering $E_8\times E_8$ heterotic string theory compactified on $\mathbb{T}^2$, for which the components in $\Gamma^{2,18}$ characterise the momentum ground states. The moduli space of such compactifications is 
described by the K\"ahler ($T$) and complex structure ($U$) moduli of ${\mathbb
T}^2$,  as well 
as by two real Wilson lines $\vec{v}_\alpha\in{\mathbb R}^{16}$, $\alpha=1,2$. At
a generic point in this 
moduli space, a general element of the momentum lattice
$\Gamma^{2,18}$ can be parametrised as $x=(m_1,n_1;m_2,n_2;\vec{\ell})$, where 
$(m_1,m_2)$ and $(n_1,n_2)$ are the momentum and winding numbers along ${\mathbb
T}^2$, while 
$\vec{\ell}\in\Lambda_{\mathfrak{e}_8} \oplus \Lambda_{\mathfrak{e}_8}$, where $\Lambda_{\mathfrak{e}_8}$ is the root lattice of the Lie algebra $\mathfrak{e}_8$. The
inner product 
on $\Gamma^{2,18}$ is defined by 
\begin{equation}\label{inpro}
\left< x|
x'\right>=-m_1n_1^{\prime}-n_1m_1^{\prime}-m_2n_2^{\prime}-n_2m_2^{\prime} 
+ \vec{\ell}\cdot \vec{\ell}^{\; \prime}\ ,
\end{equation}
where the first four terms represent the Lorentzian inner product on 
$\Gamma^{2,2}\simeq \Pi^{1,1}\oplus \Pi^{1,1}$, and the last term is the
standard 
Euclidean inner product inherited from 
${\mathbb R}^{16}\supset \Lambda_{\mathfrak{e}_8} \oplus
\Lambda_{\mathfrak{e}_8}$.
For a given vector $x=(m_1,n_1;m_2,n_2;\vec{\ell})\in \Gamma^{2,18}$, the actual internal
momentum is then 
a vector in ${\mathbb R}^{16}$:
\begin{equation}\label{bdef}
\vec{P}(x) = n_1\, \vec{v}_1 + n_2\, \vec{v}_2 + \vec{\ell} \ .
\end{equation}
In the following we want to consider the subspace of the moduli space where the
Wilson lines $\vec{v}_\alpha$ break the $\mathfrak{e}_8\oplus \mathfrak{e}_8$ gauge 
symmetry to a fixed unbroken gauge symmetry ${\mathfrak h}$, 
\begin{equation}
\mathfrak{e}_8 \oplus \mathfrak{e}_8 \quad \longrightarrow \quad {\mathfrak h} \qquad
\hbox{with}\qquad
{\mathfrak h} \oplus {\mathfrak g} \subset \mathfrak{e}_8 \oplus \mathfrak{e}_8 
\ ,
\label{DecomposeLie}
\end{equation}
where  ${\mathfrak g}$ is the maximal
commuting subalgebra in $\mathfrak{e}_8 \oplus \mathfrak{e}_8$. 
To describe this compactly, we first combine $\vec{v}_1$ and $\vec{v}_2$ into a
complex Wilson line 
$\vec{V}=\vec{v}_1 + i\vec{v}_2$. Given $\vec{V}$ we then denote by 
$\Lambda_{{\mathfrak h}}$ the sublattice of 
$\Lambda_{\mathfrak{e}_8} \oplus \Lambda_{\mathfrak{e}_8}$ 
consisting of all vectors that are orthogonal to the complex Wilson line
$\vec{V}$ (or 
equivalently to both real Wilson lines $\vec{v}_\alpha$),
\begin{equation}
\Lambda_{{\mathfrak h}} = \left\{ \vec{d} \in  \Lambda_{\mathfrak{e}_8} \oplus
\Lambda_{\mathfrak{e}_8} \ :  \
\vec{d} \cdot \vec{V} = 0\right\}  \ .
\end{equation}
The vectors of length squared two in $\Lambda_{{\mathfrak h}}$ are the roots of
the unbroken Lie algebra ${\mathfrak h}$. The commutant of
${\mathfrak h}$ in 
$\mathfrak{e}_8\oplus \mathfrak{e}_8$ defines the Lie algebra ${\mathfrak g}$,
whose root
lattice is spanned by the roots of $\mathfrak{e}_8\oplus \mathfrak{e}_8$ that
are orthogonal to
$\Lambda_{{\mathfrak h}}$ (see equation (\ref{DecomposeLie})). Adding to the corresponding root lattice the
${\mathbb T}^2$ torus directions
in $\Gamma^{2,2}$ leads to the sublattice $\Gamma_{{\mathfrak g}}\subseteq
\Gamma^{2,18}$. More formally, 
$\Gamma_{{\mathfrak g}}$ is defined as 
\begin{equation}\label{22kdef}
\Gamma_{{\mathfrak g}} = \left\{ x\in \Gamma^{2,18} \ : \ \vec{P}(x) \in
\Lambda_{{\mathfrak h}}^\perp \right\} \ ,
\end{equation}
where $\vec{P}(x)$ was defined in (\ref{bdef}). Note that the root lattice $\Lambda_{\mathfrak g}$ is the sublattice
of $\Gamma_{{\mathfrak g}}$  generated by the vectors of the form 
$(0,0;0,0;\vec{\ell})\in\Gamma_{{\mathfrak g}}$, where $\vec{\ell}$ is 
orthogonal to $\Lambda_{{\mathfrak h}}$. If $\mathfrak{g}$ is semisimple (i.e. $\mathfrak{g}=\bigoplus_{i=1}^n\mathfrak{g}_{(i)}$ with $n>1$ and $\mathfrak{g}_{(i)}$ simple Lie algebras), we will adopt the notation
\begin{align}
&\vec{\ell}=(\vec{\ell}_1,\ldots,\vec{\ell}_n)\,,&&\text{with} &&\vec{\ell}_i\in\Lambda_{\mathfrak{g}_{(i)}}\,,
\end{align}
and the inner product inherited from $\Gamma^{2,18}$ becomes 
\begin{align}
\langle x|x'\rangle&=-m_1n'_1-m'_1n_1-m_2n'_2-m'_2n_2+\sum_{i=1}^n\vec{\ell}_i\cdot \vec{\ell}^{\,\,'}_i\,, \qquad \quad x,x'\in\Gamma_{\mathfrak{g}}. \label{InnerProdGammag}
\end{align}
Furthermore, by definition of
$\Lambda_{{\mathfrak h}}$, 
$n_1 \vec{v}_1 + n_2\vec{v}_2 \in \Lambda_{{\mathfrak h}}^\perp$, and hence
$\Gamma_{{\mathfrak g}}$ has 
signature $(2,2+k)$ where $k={\rm rk}({\mathfrak g})$. Since $\Lambda_{\mathfrak
h}$ is naturally a 
sublattice of $\Gamma^{2,18}$ (corresponding to choosing $n_i=m_j=0$), we have 
\begin{equation}\label{sublat}
\Gamma_{{\mathfrak g}} \oplus \Lambda_{{\mathfrak h}}  \subseteq \Gamma^{2,18} \
,
\end{equation}
and the sublattice on the left hand side is of maximal rank.  
Generically, $\Gamma_{{\mathfrak g}} \oplus \Lambda_{{\mathfrak h}}$ is a 
proper sublattice of $\Gamma^{2,18}$ with index $s$. We can write the
decomposition
\begin{equation}\label{latdec}
\Gamma^{2,18} = (\Gamma_{{\mathfrak g}} \oplus \Lambda_{{\mathfrak h}} )\ 
\oplus \ \bigoplus_{\mu=1}^{s-1} 
\left( \lambda_\mu + \Gamma_{{\mathfrak g}} \oplus \Lambda_{{\mathfrak h}} 
\right) \ ,
\end{equation}
where $\lambda_\mu$ denotes the different cosets. More precisely, this construction can be understood as follows. While the lattice
$\Gamma_{{\mathfrak g}} \oplus \Lambda_{{\mathfrak h}}$ is integral Euclidean it
is in general not self-dual and the (finite) quotient group $(\Gamma_{{\mathfrak
g}} \oplus \Lambda_{{\mathfrak h}})^*/(\Gamma_{{\mathfrak g}} \oplus
\Lambda_{{\mathfrak h}})$ is typically non-trivial. We therefore choose a set of
generators $\lambda_\mu$ to write the coset representatives of the glue group as
(see e.g.~\cite{Lust:1989tj,Gannon:1991vg})
\begin{align}
&(\Gamma_{{\mathfrak g}} \oplus \Lambda_{{\mathfrak h}})^*/(\Gamma_{{\mathfrak
g}} \oplus \Lambda_{{\mathfrak
h}})=\left\{\lambda_\mu\right\}\,,&&\text{with}&&\lambda_\mu\in\Gamma_{{
\mathfrak g}} \oplus \Lambda_{{\mathfrak h}} \,.
\label{cosetgroup}
\end{align}
Here the conjugacy classes $\mu$ are called the \emph{glue classes} and
$\lambda_\mu$ is sometimes referred to as the glue vector. The union of all glue
vectors in all glue classes $\mu$ forms the integral lattice $\Gamma^{2,18}$.
The order of the glue group (i.e. the number of different glue classes
$\mu=0,\ldots, s-1$) is given by $|\Lambda_{\mathfrak{g}}|=s$. 
In some cases we also introduce
$\lambda_0\equiv 0$, and include $\mu=0$ in (\ref{latdec}).

In order to get an intuition for these various sublattices it is useful to
consider the `extremal' 
cases.  For generic Wilson line moduli $\vec{V}$, then $\Lambda_{{\mathfrak h}} = \{0\}$,
and the condition for 
$\vec{P}(x)$ to be orthogonal to $\Lambda_{{\mathfrak h}}$ is empty; in this
case, 
${\mathfrak g} = {\mathfrak e}_8 \oplus {\mathfrak e}_8$, and the lattice
$\Gamma_{{\mathfrak g}}$ has signature $(2,18)$. This is the case discussed at length in \cite{GHPI}. The other extremal case arises
if $\vec{V}=\vec{0}$, in which case
$\Lambda_{{\mathfrak h}} = \Lambda_{\mathfrak{e}_8} \oplus 
\Lambda_{\mathfrak{e}_8}$. 
Then the condition to be orthogonal to $\Lambda_{{\mathfrak h}}$ means that
$\vec{P}(x)=0$, 
and $\Gamma_{{\mathfrak g}} \cong \Pi^{2,2}$ has signature $(2,2)$ and is
generated by $n_i$ and $m_j$. 
For suitable intermediate choices of Wilson lines, however, we can also get
lattices that
lie in between. 

\subsection{The Wilson Line Moduli}\label{Sect:WilsonLine}

As is clear from this discussion the lattice decomposition (\ref{sublat})
depends on the choice of 
Wilson lines. In the following we want to study the submanifold of the moduli
space (which we shall
call ${\cal M}_{2,2+k}$, where $k$ is the rank of ${\mathfrak g}$) along which
this 
decomposition is constant. One way to guarantee this is to fix the `direction'
of $\vec{V}$ in the following 
way. Let us introduce a basis $\vec{e}_i$, $i=1,\ldots,16$ for  
${\mathbb R}^{16}$ consisting of the simple roots of $\mathfrak{e}_8\oplus
{\mathfrak e}_8$. 
 (Thus, in particular, $\vec{e}_i \cdot \vec{e}_j = \mathcal{C}_{ij}$, with 
$\mathcal{C}_{jk}$ the  Cartan matrix of $\mathfrak{e}_8 \oplus {\mathfrak
e}_8$.) 
We denote the dual basis by 
$\vec{f}^{\, l}$, $l=1,\ldots, 16$ so that  $\vec{e}_j \cdot \vec{f}^{\, l} =
\delta_j^l$, and write 
$\vec{V}$ in this basis,  {\it i.e.}\  $\vec{V} = V_j \, \vec{f}^{\, j}$. If
$\vec{V}$ has precisely 
$k$ non-zero coefficients $V_j$, say $V_{l(1)},\ldots, V_{l(k)}$, then the 
decomposition (\ref{sublat}) is generically independent of the actual values of 
these coefficients. Indeed, $\Lambda_{{\mathfrak h}}$ is then a
$(16-k)$-dimensional lattice generated by 
\begin{equation}
\Lambda_{{\mathfrak h}} = \text{span}_{\mathbb{Z}} \Bigl( e_j \   \big| \ j \not\in
\{l(1),\ldots,l(k) \} \Bigr) \ .
\end{equation}
We parametrise an arbitrary point in this  moduli space ${\cal M}_{2,2+k}$ by
\be
\y=(U,T; \vec{V})\in {\mathbb C}^{1,1+k},
\ee 
where $\vec{V}=(V_1,\ldots,V_k)$ and $V_j$ is the component with respect to
$\vec{f}^{\, l(j)}$, $j=1,\ldots, k$. 
On the space ${\mathbb C}^{1,1+k}$ we have the inner product
$(\y|\y')=-TU'-T'U+\vec{V}\cdot \vec{V}'$ such that
\begin{equation}
(\y|\y) = - 2\, T\, U + \vec{V}^2 \ .\label{innerProdRed}
\end{equation}
For the following it is also useful to define the map (see \cite{Harvey:1995fq})
\begin{equation}
u : {\mathbb C}^{1,1+k} \rightarrow {\mathbb C}^{2,2+k} \ , \quad 
\y=(U,T; \vec{V}) \ \mapsto \ u(\y) = \left(U,T;   \frac{(\y|\y)}{2},1 ;
\vec{V}\right) \ ,
\end{equation}
which associates to every element $\y\in {\mathbb C}^{1,1+k}$ a light-like
vector 
$u(\y)\in {\mathbb C}^{2,2+k}$. Here the inner product on ${\mathbb C}^{2,2+k}$
is defined
by $\langle \cdot | \cdot \rangle$ as in (\ref{InnerProdGammag}). With this notation an
arbitrary momentum state $x\in\Gamma_{\mathfrak g}$ parametrised by 
$x=(m_1,n_1;m_2,n_2;{\vec{\ell}}\,\,)$ has $|p^L|^2 = -2 |\left< x | u(\y) \right>|^2/Y$, where $Y=(\Im \y|\Im \y)$ and $\Im \y=(U_2,T_2, \Im \vec{V})$ is the imaginary part of $\y$. We further have $\left< x | x\right> =\bigl(|p^{R}|^2-|p^{L}|^2\bigr)$.

%%%%%%%%%%%%%%%%%%%%%%%%%%%%%%%%%%%%%%%%%%%%%%%%%%%%%%%%%%%%%%%%%%%%%
%%%%%%%%%%%%%%%%%%%%%%%%%%%%%%%%%%%%%%%%%%%%%%%%%%%%%%%%%%%%%%
\section{Topological Amplitudes and Denominator Formulas}\label{Section:Akintegral}
\setcounter{equation}{0}

In this section we introduce and analyse a particular class of topological
$\mathcal{N}=4$ amplitudes $\mathcal{F}_g$ in heterotic string theory compactified on 
$\mathbb{T}^6$ (see \cite{Antoniadis:2006mr,Antoniadis:2007cw}). We evaluate the one-loop integral corresponding to a particular analytic part $\mathcal{F}_{1}^{\text{analy}}$ of the amplitude, for a generic unbroken gauge algebra $\mathfrak{h}$. We show that part of the result can be identified with the infinite product side of the denominator formula for a certain Borcherds extension of $\mathfrak{g}^{++}$ (for details on Borcherds algebras see Appendix \ref{App:BorcherdsExtension}). To verify the automorphic properties of the denominator formula we find an explicit integral representation of (the logarithm of) the infinite product, which corresponds to a theta correspondence for congruence subgroups of $SO(2,2+k;\mathbb{Z})\times SL(2,\mathbb{Z})$. 
\subsection{$\cN=4$ Topological Amplitudes}
\label{N4Top}
In the naive field-theory limit the couplings $\mathcal{F}_g$  only receive contributions 
from perturbative $1/2$ BPS-states. However, in string theory additional  
non-analytic terms appear as well. We shall make use of the key observation from our previous work \cite{GHPI}, namely that 
one may use the harmonicity equations satisfied by $\mathcal{F}_g$ to isolate an analytic 
part $\mathcal{F}_g^{\text{analy}}$. In a sense, this represents the $\mathcal{N}=4$ analogue
of the `threshold corrections' 
in $\mathcal{N}=2$ theories. 

In \cite{Antoniadis:2006mr,Antoniadis:2007cw} (see also \cite{Antoniadis:2007ta}) a particular class of $\cN=4$ topological string amplitudes has been discovered. These amplitudes appear at the $g$-loop level in type~II string theory compactified on $K3\times {\mathbb T}^2$, while their dual counterparts in heterotic string theory compactified on ${\mathbb T}^6$ start receiving contributions at the one-loop level. We will focus on the case $g=1$ for which the latter amplitude takes the following form 
\begin{align}
\mathcal{F}_{1}(\y)=\int_{\mathbb{F}}\f{d^2\tau}{\bar{\eta}^{24}}\,\tau_2\,G_{2}(\tau,\bar{\tau})\Theta^{(6,22)}(\tau,\bar{\tau},\y)\ , \label{GenTopAmplitude}
\end{align}
where the integral is over the fundamental domain $\mathbb{F}:=\mathbb{F}(\Gamma)=\mathbb{H}/\Gamma$ of $\Gamma=SL(2,\mathbb{Z})$, where $\mathbb{H}$ is the standard upper half plane. Moreover, the expression
\begin{equation}
\Theta^{(6,22)}(\tau,\bar{\tau},\y)=\sum_{p\in\Gamma^{6,22}\atop  p\neq 0}\,q^{\frac{1}{2}|p^L|^2}\,\bar{q}^{\frac{1}{2}|p^R|^2}\,,\label{GeneralNarainTheta}
\end{equation}
is a Siegel-Narain theta-function (without momentum insertions) of the even unimodular lattice $\Gamma^{6,22}$. Notice that we do not sum over $p=0$ in the definition of $\Theta^{(6,22)}(\tau,\bar{\tau},\y)$. As was explained in \cite{GHPI} this is a particular choice of regularization which removes an overall singularity of $\mathcal{F}_1$. The object $G_{2}(\tau,\bar{\tau})$ in (\ref{GenTopAmplitude}) is a weight $4$ non-antiholomorphic modular form. The explicit expression was computed in \cite{Antoniadis:1995zn} (see also \cite{Antoniadis:2010iq}) and is given by
\begin{align}
G_{2}(\tau,\bar{\tau})=\zeta(4)\left(\bar{E}_4(\tau)+5\hat{\bar{E}}^2_2(\tau,\bar{\tau})\right)\,,&&\text{with} &&\hat{\bar{E}}_2(\tau, \bar\tau)= E_2(\bar{\tau})-\f{3}{\pi \tau_2}\ ,
\end{align}
where $E_{2k}(\tau)$ is the weight $2k$ Eisenstein series. Notice that $E_2$ is a 'quasi-modular form' \cite{Dijkgraaf,KanekoZagier}, which means that in addition to a weight factor it also receives an anomalous shift-term under modular transformations. Therefore, following standard practice, we have introduced the quantity $\hat{\bar{E}}_2$ which is an honest weight 2 modular form, but non-antiholomorphic in $\tau$. It is natural to decompose $G_{2}(\tau,\bar{\tau})$ into an analytic (antiholomorphic) and non-analytic part
\begin{align}
&G_2(\tau,\bar{\tau})=G_2^{\text{analy}}(\bar{\tau})+G_2^{\text{non-analy}}(\tau,\bar{\tau})\,,&&\text{with} &&\begin{array}{l}G_2^{\text{analy}}(\bar{\tau})=\zeta(4)\bar{E}_4(\tau)\,, \\[6pt] G_2^{\text{non-analy}}(\tau,\bar{\tau})=5\zeta(4)\hat{\bar{E}}^2_2(\tau,\bar{\tau})\,.\end{array} \label{BPSgfunct}
\end{align}
In \cite{GHPI} this splitting was proposed based on the fact that (for generic $g$) the non-analytic part is responsible for an anomalous violation of particular supersymmetric Ward-identities \cite{Antoniadis:2007cw} (``harmonicity relations``) satisfied by the amplitudes $\mathcal{F}_g$ at the string quantum level. 

At particular points where the gauge group is enhanced due to the presence of additional massless bosons, some additional care is needed since the harmonicity relations require regularization of certain singular contributions. However, once this subtlety has been properly addressed, we will continue using the definition of $G_g^{\text{analy}}(\bar{\tau})$ and discard the remaining non-antiholomorphic terms. 

As the main object of study, we thus introduce the analytic one-loop integral
\begin{align}
\mathcal{F}_{1}^{\text{analy}}({\y})=\int_{\mathbb{F}}\f{d^2\tau}{\bar{\eta}^{24}} \tau_2\,G_{2}^{\text{analy}}(\bar{\tau})\Theta^{(6,22)}(\tau,\bar{\tau},\y)\ .\label{BPSamplitudeDef}
\end{align}
We recall that the decomposition (\ref{BPSgfunct}) does not break modular invariance and therefore the integral $\mathcal{F}_{1}^{\text{analy}}(\y)$ is well defined. 

As in Section \ref{NarainComp} we will consider the 
internal six-torus to be factorised as ${\mathbb T}^6={\mathbb T}^4\times
{\mathbb T}^2$, 
and take the large volume limit of ${\mathbb T}^4$. This implies that the Siegel-Narain theta function 
of the original $\Gamma^{6,22}$ Narain-lattice decomposes according to
\begin{align}
\frac{G_{2}^{\text{analy}}(\bar{\tau})\tau_2^2}{\bar{\eta}^{24}}\,\Theta^{(6,22)}
\sim\text{Vol}\,\frac{G_{2}^{\text{analy}}(\bar{\tau})}{\bar{\eta}^{24}}\,\Theta^{(2,18)} \ ,
\label{PreReductionSiegelNarain}
\end{align}
where $\text{Vol}$ is the volume of ${\mathbb T}^4$ and $\Theta^{(2,18)}$ the Siegel-Narain theta function of the lattice $\Gamma^{2,18}$ appearing in (\ref{LatticeSplitLargeVolume}). As we shall see in section~\ref{Section:Akintegral}, due to the choice of Wilson line $\vec{V}$ described in section~\ref{Sect:WilsonLine}, the $\Gamma^{2,18}$ lattice will be decomposed even further. For the time being, however, we will study more closely (\ref{PreReductionSiegelNarain}) which will already teach us some valuable lessons about the algebraic properties of $\mathcal{F}_1^{\text{analy}}$.

For example, one can show that the integral $\mathcal{F}_{1}^{\text{analy}}$ develops singularities at complex codimension one submanifolds of $SO(2,2+k)/(SO(2)\times SO(k))$, which coincide with the walls of the (complexified) fundamental Weyl chamber of the hyperbolic extension $\mathfrak{g}^{++}$ of the broken part $\mathfrak{g}$ of the gauge algebra $\mathfrak{e}_8\oplus \mathfrak{e}_8$. As a consequence, the singularity behaviour of the BPS-spectrum is controlled by the hyperbolic Weyl group $\mathcal{W}(\mathfrak{g}^{++})$, similarly as for the non-perturbative $1/4$ BPS dyon spectrum \cite{Cheng:2008fc}. In order to better understand the underlying algebraic structure, we will now proceed to evaluate the integral $\mathcal{F}_{1}^{\text{analy}}$ explicitly. 
%%%%%%%%%%%%%%%%%%%%%%%%%%%%%%%%%%%%%%%%%%%%%%%%%%%%%%%
%%%%%%%%%%%%%%%%%%%%%%%%%%%%%%%%%%%%%%%%%%%%%%%%%%%%%%%%%%%%%%%%%%%%%%%%%%%%%%%%
\subsection{One-Loop Integral for any Choice of Gauge Group}\label{Sect:TorusIntegralEval}
The first step to explicitly perform the analytic one-loop integral (\ref{PreReductionSiegelNarain}) for arbitrary choices of the gauge group is to implement the splitting (\ref{DecomposeLie}) at the level of the Siegel-Narain theta function $\Theta^{(2,18)}(\tau,\bar{\tau},\y)$. Since we are interested in the submanifold ${\cal M}_{2,2+k}$ of the
moduli space along which only some components of $\vec{V}$ are non-zero, we can use the same decomposition as in (\ref{latdec}) of the lattice $\Gamma^{2,18}$ and write 
\begin{align}
\frac{G_{2}^{\text{analy}}(\bar{\tau})}{\bar{\eta}^{24}}\,\Theta^{(2,18)}(\tau,\bar{\tau},\y)
\equiv \sum_{\mu=0}^{s-1}\mathcal{P}^{(k)}_\mu(\bar{\tau})\,
\Theta^{(2,2+k)}_{\mu}(\tau,\bar{\tau},\y)\ .
\label{ReductionSiegelNarain}
\end{align}
Here $\Theta^{(2,2+k)}_{\mu}(\tau,\bar{\tau},\y)$ is the theta function associated to the $\Gamma_{\mathfrak g}$ coset $\lambda_\mu$ (see (\ref{latdec})),
\begin{align}
&\Theta^{(2,2+k)}_{\mu}(\y)
=\sum_{x \in\Gamma_{{\mathfrak g}}+\lambda_{\mu}^{\mathfrak g}}\bar{q}^{\frac{1}{2}\,\langle x|x\rangle}\,
e^{2\pi \tau_2\frac{|\langle x|u(\y)\rangle|^2}{(\Im \y|\Im
\y)^2}}\,,\label{SiegelNarain22k}
\end{align}
while $\mathcal{P}^{(k)}_\mu(\bar{\tau})$ captures the contributions from $G_{2}^{\text{analy}}/\bar{\eta}^{24}$ and the theta constants from the different $\Lambda_{\mathfrak{h}}$ cosets
\begin{align}
&\mathcal{P}_\mu^{(k)}(\bar{\tau})
=\frac{G_2^{\text{analy}}}{\bar{\eta}^{24}}\,
\Theta_\mu^{\mathfrak{h}}(\bar{\tau})=\frac{G_2^{\text{analy}} 
}{\bar{\eta}^{24}}\,
\sum_{\vec{\ell}\in\Lambda_{\mathfrak{h}}+\lambda_\mu^{\mathfrak
h}}\bar{q}^{\frac{1}{2}\,\vec{\ell}\cdot\vec{\ell}}\,.
\label{Pmuexpression}
\end{align}
Since  $\Lambda_{{\mathfrak h}}$ is a sublattice of $\mathfrak{e}_8\oplus \mathfrak{e}_8$ that is orthogonal to $\vec{V}$, $\mathcal{P}^{(k)}_\mu(\bar{\tau})$ does not depend on the moduli $\y=(U,T;\vec{V})$. In (\ref{SiegelNarain22k}) and (\ref{Pmuexpression}) $\mu$ labels the $s$ conjugacy classes of the lattices as in (\ref{latdec}), while $\lambda_\mu^{\mathfrak g}$ and  $\lambda_\mu^{\mathfrak h}$ are the projections of $\lambda_\mu$ onto ${\mathfrak g}$ and ${\mathfrak h}$, respectively. We will parametrise the summation in (\ref{SiegelNarain22k}) by $x = (m_1,n_1;m_2,n_2;\vec{\ell}\,\,)$ with $\vec{\ell}\in \Lambda_{\mathfrak{g}}+\lambda_\mu^{\mathfrak g}$, and similarly for (\ref{Pmuexpression}).

Putting things together, the analytic integral can be written in the following form
\be 
\mathcal{F}_1^{\text{analy}}(\y)=\int_{\mathbb{F}} d^2\tau\, \tau_2  \sum_{\mu=0}^{s-1}\mathcal{P}^{(k)}_\mu(\bar{\tau})\,
\Theta^{(2,2+k)}_{\mu}(\tau,\bar{\tau},\y).
\label{F1thetalift}
\ee
An integral of this type has already been computed in \cite{GHPI} by splitting the integral in different orbits with respect to $SL(2,\mathbb{Z})$ (this method was first developed in \cite{Dixon:1990pc} and 
further extended in \cite{Harvey:1995fq,LopesCardoso:1996nc,Kiritsis:1997hf,Lerche:1998nx,Foerger:1998kw,Obers:1999um}).
%\begin{align}
%\mathcal{F}_1^{\text{analy}}(\y)=\mathcal{I}_0^{\text{(analy)}}+\mathcal{I}_{\text{ND}}^{\text{(analy)}}+\mathcal{I}_{\text{D}}^{\text{(analy)}}\,.
%\label{zeroDND}
%\end{align}
Generalizing the result of \cite{GHPI} to the case of arbitrary gauge groups we arrive at the following explicit expression 
\begin{align}
\mathcal{F}_{1}^{\text{analy}}(\y)
=\sum_{\mu=0}^{s-1}&\bigg\{\sum_{{\vec{\ell}\in\Lambda_{\mathfrak{g}}+\lambda_\mu^{\mathfrak g}}}
\bigg[\frac{2\pi
Y}{3U_2}\left(c_{\mu}(0,\vec{\ell})-24c_{\mu}(-1,\vec{\ell}
)\right)+2\log\left|1-e^{2\pi i\vec{\ell}\odot\vec{V}}\right|^{c_{\mu}(0,\vec{\ell})} \nonumber \\
&+2\log\prod_{{n',r\in \mathbb{Z}}\atop r>0}
\left|1-e^{2\pi i(rT+n'U+\vec{\ell}\odot\vec{V})}\right|^{c_{\mu}(n'r,\vec{\ell})}+2\log\prod_
{n=1}^\infty\left|1-e^{2\pi
i(nU+\vec{\ell}\odot{\cdot}\vec{V})}\right|^{c_{\mu}(0,\vec{\ell})}\bigg]
\nonumber \\
& 
+c_{\mu}(0,\vec{0})\left(\frac{\pi U_2}{3}-\ln Y+K\right)
+2\log\prod_{n=1}^\infty|1-e^{2\pi in U}|^{c_{\mu}(0,0)}\bigg\} \nonumber \\
&\hspace*{-0.8cm} 
+\frac{2U_2}{3\pi}+\frac{2\pi}{U_2}(\vec{\ell}\odot\Im\vec{V})\left((\vec{\ell}
\odot\Im\vec{V})+U_2\right)\,,\label{TorusIntegralFinResult}
\end{align}
where $K=\gamma_E-1-\ln\frac{8\pi }{3\sqrt{3}}$, with $\gamma_E$ being the Euler-Mascheroni constant. Furthermore, we have introduced the shorthand notation for the modified scalar-product: $\vec{\ell}\odot\vec{V}=\ell\cdot\Re\vec{V}+i |\vec{\ell}\cdot\Im\vec{V}|\,$. The coefficients $c_\mu(n'r, \vec{\ell})$ arise from the Fourier expansion
\begin{align}
\sum_{\mu=0}^{s-1}\mathcal{P}_\mu^{(k)}(\bar{\tau})
\sum_{\vec{\ell}\in\Lambda_{\mathfrak{g}}+\lambda_\mu^{\mathfrak g}}\,
\bar{q}^{\frac{1}{2}\vec{\ell}\cdot \vec{\ell}}\,
e^{2\pi i\vec{\ell}\cdot \vec{z}}=\sum_{\mu=0}^{s-1}\sum_{n= -1}^\infty
\sum_{\vec{\ell}\in\Lambda_{\mathfrak{g}}+\lambda_\mu^{\mathfrak g}} 
c_{\mu}(n,\vec{\ell}\,\,)\,\bar{q}^n\,e^{2\pi i \vec{\ell}\cdot
\vec{z}}\,.\label{FourierExpansionConjugacy}
\end{align}
For some simple examples, explicit expressions for $c_{\mu}(n,\vec{\ell}\,\,)$ will be given in section~\ref{Sect:Examples}. 

By construction, (\ref{FourierExpansionConjugacy}) transforms as a weak Jacobi form under $SL(2,\mathbb{Z})$, and thus the coefficients $c_{\mu}(n,\vec{\ell}\,\,)$ only depend on $(n,\vec{\ell}\,\,)$ through the combination $(n-\tfrac{1}{2}\vec{\ell}\cdot\vec{\ell}\, )$ \cite{EZ}. Moreover, by inspection of (\ref{Pmuexpression}) it is clear that the integrand in (\ref{F1thetalift}) has a simple pole at $\tau\to i\infty$,  hence
\begin{equation}
c_{\mu}\left(n-\tfrac{1}{2}\vec{\ell}\cdot\vec{\ell}\right)=0 \qquad \forall\,
n-\tfrac{1}{2}\vec{\ell}\cdot\vec{\ell}<-1\ .\label{CondFourCoeffVan}
\end{equation}
In the following we shall mainly be interested in the contribution of the trivial conjugacy class labelled by $\mu=0$. For this the only terms in the sum over $\vec{\ell}$ with $\vec{\ell}\neq \vec{0}$ come from the degenerate orbit and have $\vec{\ell}\cdot\vec{\ell}=2$. We can then choose to work in a chamber of the moduli space where $ \Im\vec{V}\in \Lambda_{\mathfrak{g}}^+\otimes \mathbb{C}\ $, for which the condition $\vec{\ell}\cdot (\Im\vec{V})>0$ can be equivalently written as $\vec{\ell}\in \Lambda_{\mathfrak{g}}^+$.

%%%%%%%%%%%%%%%%%%%%%%%%%%%%%%%%%%%%%%%%%%%%%%%%%%%%%%%%%%%%%%%%%%%%%%%%%%%%%%%%%%%%%%%%%%%
\subsection{Borcherds Lift and Denominator Formula}\label{Sect:DenominatorFormula}
In the following we shall restrict the analysis to a particular part of the full analytic amplitude (\ref{TorusIntegralFinResult}), corresponding to the contribution of the trivial conjugacy class $\mu=0$. We will show that this may be identified with the  infinite product side of the denominator formula for the Borcherds extension $ \bor(\mathfrak{g}^{++})$, where $\mathfrak{g}^{++}$ is the double extension of the unbroken gauge algebra $\mathfrak{g}$.\footnote{For details on double extensions of Lie algebras in this context see Appendix A of \cite{GHPI}.} From this point of view the reason for restricting to $\mu=0$ becomes clear: only for the zero conjugacy class does the sum over $\vec{\ell}$ correspond to a sum over roots of $\mathfrak{g}$. Indeed, the higher conjugacy classes give rise to sums over weights of $\mathfrak{g}$ rather than roots.

%%%%%%%%%%%%%%%%%%%%%%%%%%%%%%%%%%%%%%%%%%%%%%%%%%%%%%%%%%%%%%%%%%%%%%%%%%%%%%%%%%%%%%
\subsubsection{Automorphic Product}
The zero conjugacy class contribution to (\ref{TorusIntegralFinResult}) can be written as
\begin{align}
\mathcal{F}_1^{\text{analy}}(\y)\big|_{\mu=0}=\log||\Phi_{\mathfrak{g}}(\y)||^2 +c_{0}(0,\vec{0})\left(\frac{\pi U_2}{3}-\ln Y+K\right)+\ldots\ ,
\label{LogTermInteg}
\end{align}
where we have defined
\be
\Phi_{\mathfrak{g}}(\y)=e^{-2\pi i (\rho|\y)}\prod_{(r,n';\vec{\ell})>0} \left(1-e^{2\pi i
(rT+n'U+\vec{\ell}\cdot\vec{V})}\right)^{c_{0}(n'r,\vec{\ell}\,\,)},
\label{Phidef}
\ee
and the norm $|| \cdot ||$ in (\ref{LogTermInteg}) takes into account the contribution with $(r,n';\vec{\ell})<0$. We will give a more precise description of this norm below, once we have analysed the modular properties of $\Phi_{\mathfrak{g}}(\y)$ in detail. For the moment we just remark that the exact range of $(r,n';\vec{\ell})$ needs to be discussed separately for the cases when $\mathfrak{g}$ is simple or semisimple. Since this discussion is mostly technical and somewhat tedious, we have relegated it to appendices \ref{simple} and \ref{semisimple} respectively. There we show that the product can be written to range over the elements $\alpha\in\Lambda_{{\mathfrak g}^{++}}^+$ with norm
$\alpha^2\leq 2$. Because of this we can then write $\Phi_{\mathfrak{g}}(\y)$ as the following product
\begin{align}
\Phi_{\mathfrak{g}}(\y)=e^{-2\pi i (\rho|\y)}\prod_{\alpha \in\Lambda_{{\mathfrak g}^{++}}^+}\,
 \left(1-e^{2\pi i (\alpha|\y)}\right)^{c_{0}(-\alpha^2/2)}\
,\label{DenomFormPosRootwoWeyl}
\end{align}
where we used $c_{0}(n)=0$ for $n<-1$. The idea is now to identify $\Phi_{\mathfrak{g}}(\y)$ with the denominator formula (\ref{denominatorformula}) for a BKM-algebra which we shall call $\bor(\mathfrak{g}^{++})$ to indicate that it is an `automorphic correction` (in the terminology of \cite{Gritsenko:1996ax}) of $\mathfrak{g}^{++}$. Indeed, we would identify $\rho$ as the Weyl vector of $\bor(\mathfrak{g}^{++})$ (see Appendix \ref{App:BorcherdsExtension}) and the multiplicities of all (real and imaginary) roots of $\bor(\mathfrak{g}^{++})$ could then be conveniently read off as the Fourier coefficients $c_{0}(n,\vec{\ell}\,\,)$ as defined by the seed function 
\begin{align}
\psi_{\mathfrak{g}}(\bar{\tau},\vec{z})=\sum_{n=-1}^\infty \sum_{\vec{\ell}\in\Lambda_{\mathfrak{g}}} c_{0}\left(n-\tfrac{1}{2}\vec{\ell}\cdot\vec{\ell}\, \right)\,\bar{q}^n\,e^{2\pi i \vec{\ell}\cdot \vec{z}}\, , \label{Jacobi}
\end{align} 
with Fourier coefficients arising from the zeroth conjugacy class $\mu =0$ in (\ref{FourierExpansionConjugacy}). However, to justify the identification of $\Phi_{\mathfrak{g}}(\y)$ with a denominator formula for $\bor(\mathfrak{g}^{++})$ we must show that $\Phi_{\mathfrak{g}}(\y)$ extends to an automorphic form on 
\begin{align}
\grg\backslash \mathcal{M}_{2,2+k}=\grg  \backslash SO(2,2+k)/(SO(2)\times SO(2+k)\,,
\end{align} 
for some discrete subgroup $\grg\subset SO(2,2+k)$. To show this we note that although the seed function $\psi_{\mathfrak{g}}(\tau, \vec{z})$ in (\ref{Jacobi}) no longer transforms nicely under the full mapping class group $\Gamma$ of the original string worldsheet  torus, it is nevertheless a weak Jacobi form with respect to a congruence subgroup $\gr{0}\subset \Gamma$.\footnote{This is in fact true for every individual $\mu$ in the right hand side of (\ref{ReductionSiegelNarain}): each summand is a weak Jacobi form of zero weight under a particular congruence subgroup $\gr{\mu}\subset \Gamma$. We notice in particular, that every single summand is invariant under the generator $T\in SL(2,\mathbb{Z})$ which acts as $T:\,\,\tau\mapsto \tau+1$.} Realising moreover that (\ref{F1thetalift}) has structurally the form of a `multiplicative' (Borcherds) lift, one might suspect that the modular properties of $\psi_{\mathfrak{g}}(\bar{\tau},\vec{z})$ with respect to $\Gamma_{[0]}$ directly translate into modular properties of $\Phi_{\mathfrak{g}}(\y)$ with respect to some subgroup $\grg$ of the T-duality group $SO(2,2+k;\mathbb{Z})$. We shall now verify that this is indeed the case.

%%%%%%%%%%%%%%%%%%%%%%%%%%%%%%%%%%%%%%%%%%%%%%%%%%%%%%%%%%%%%%%%%%%%%%
\subsubsection{Theta Correspondence and Modular Properties}
As already mentioned, the integral representation (\ref{F1thetalift}) of the amplitude $\mathcal{F}_1^{\text{analy}}$ provides an example of a so called \emph{theta correspondence}. Since this notion will play an important role in what follows, we begin this section with a brief review of the key features (see, e.g, \cite{Borcherds2,Kontsevich,Prasad} for more details). 

Let $(G_1, G_2)$Ê be a dual reductive pair of Lie groups in the sense of Howe \cite{Howe}. This means that the product $G_1\times G_2$ is a subgroup of (the universal cover of ) a symplectic group $Sp(W)$, with $W$ a symplectic vector space, such that $G_1$ (resp. $G_2$) is the centraliser of $G_2$ (resp. $G_1$) inside $Sp(W)$. The standard example is when $G_1=SL(2,\mathbb{R})$ and $G_2=SO(m,n)$ such that $SL(2,\mathbb{R})\times SO(m,n)\subset Sp(2(m+n))$. Automorphic forms correspond to irreducible components in the decomposition of $L^2\big(G_1(\mathbb{Z})\backslash G_1\big)$ and $L^2\big(G_2(\mathbb{Z})\backslash G_2\big)$, where $(G_1(\mathbb{Z}), G_2(\mathbb{Z}))$ are discrete subgroups. In a nutshell, the theta correspondence is then an integral transform from automorphic representations of $G_1$ to automorphic representations of $G_2$. For the reductive pair $(SL(2,\mathbb{R}), SO(m,n))$, the kernel of this integral transform is a Siegel-Narain theta series $\Theta^{(m,n)}(\tau, \bar{\tau}, \y)$, as exemplified by (\ref{GenTopAmplitude}) for $(m,n)=(6,22)$.

The purpose of this section is to determine the modular properties of the infinite product $\Phi_{\mathfrak{g}}(\y)$ in (\ref{DenomFormPosRootwoWeyl}). We shall do this by utilizing the theta correspondence outlined above. We thus seek an integral transform from a $ \gr{0} \subset SL(2,\mathbb{Z})$ modular form to an automorphic form for $\grg  \subset SO(2,2+k;\mathbb{Z})$ which can be identified with $\Phi_{\mathfrak{g}}(\y)$. To this end, we assume that $\gr{0}$ has finite index $N$ (we will see that this is indeed the case in all examples discussed in section~\ref{Sect:Examples}), for which we can choose coset representatives $\gamma_1,\ldots, \gamma_N$ such that $\Gamma$ can be written as the disjoint union (see e.g.~\cite{Gunning})
\begin{align}
\Gamma=\gamma_1\gr{0}\cup\ldots \cup  \gamma_N\gr{0}\,.
\end{align}
We can then construct a fundamental domain of $\gr{0}$ by
\begin{align}
\mathbb{F}_{[0]}:=\gamma_1\mathbb{F}\cup\ldots\cup \gamma_N\mathbb{F}=\mathbb{H}/\gr{0}\,,
\end{align}
where $\mathbb{F}=\mathbb{H}/\Gamma$ is the fundamental domain of $\Gamma$. Using this result we can rewrite $\mathcal{F}_1^{\text{analy}}$ in the following manner
\begin{align}
\mathcal{F}_1^{\text{analy}}(\y)=&\phantom{+}\frac{1}{N}\int_{\mathbb{F}_{[0]}} d^2\tau\, \tau_2  \mathcal{P}^{(k)}_0(\bar{\tau})\,
\Theta^{(2,2+k)}_{0}(\tau,\bar{\tau},\y)\nonumber\\
&+\frac{1}{N}\int_{\mathbb{F}_{[0]}} d^2\tau\, \tau_2  \sum_{\mu=1}^{s-1}\mathcal{P}^{(k)}_\mu(\bar{\tau})\,
\Theta^{(2,2+k)}_{\mu}(\tau,\bar{\tau},\y)\,.\label{DenomSplitConsis}
\end{align}
Notice that this is indeed a consistent splitting of the integral, since the integrands of both terms separately are modular invariant under $\gr{0}$ in the fundamental domain $\mathbb{F}_{[0]}$. We can now use similar methods as developed in \cite{Dixon:1990pc} (and 
further extended in \cite{Harvey:1995fq,LopesCardoso:1996nc,Foerger:1998kw,Lerche:1998nx,Antoniadis:2009tr,Obers:1999um})\footnote{See also e.g. \cite{FunkeBruinier} for a recent treatment of such integrals in the mathematics literature.}  to evaluate the first term of (\ref{DenomSplitConsis}) separately. To this end we first perform a Poisson resummation to obtain
\begin{align}
&\int_{\mathbb{F}_{[0]}} d^2\tau\, \tau_2  \mathcal{P}^{(k)}_0(\bar{\tau})\,
\Theta^{(2,2+k)}_{0}(\tau,\bar{\tau},\y)=\int_{\mathbb{F}_{[0]}}\frac{d^2\tau}{\tau_2^2}
\frac{Y}{U_2}\,\mathcal{P}_0^{(k)}(\bar{\tau})\sum_{{(p_1,n_1;p_2,n_2)}\atop\vec{\ell}\in \Gamma_{\mathfrak{g}}}\,
\bar{q}^{\frac{1}{2}\vec{\ell}\cdot \vec{\ell}}\,e^{F(A,\y)}\,  \label{IntegralbeforeOrbits}
\end{align}
with the shorthand notation
\begin{align}
F(A,\y)=2\pi i\vec{\ell}\cdot
\vec{z}-\frac{\pi Y}
{U_2^2\tau_2}|\mathcal{A}|^2-2\pi iT\text{det}A-\frac{\pi n_2
\left(\vec{V}^2\tilde{\mathcal{A}}-\vec{\bar{V}}^2\mathcal{A}\right)}{U_2}
+\frac{2\pi i\,(\Im\vec{V})^2}{U_2^2}(n_1+n_2\bar{U})\mathcal{A}\nonumber
\end{align}
where $p_1,p_2,n_1,n_2\in\mathbb{Z}$ such that $(p_1,n_1;p_2,n_2;\vec{\ell})\neq (0,0;0,0;\vec{0})$ and the matrices $(A, \mathcal{A}, \tilde{\mathcal{A}})$ are the same as in \cite{GHPI}. We have also used the shorthand expression $\vec{z}=\tfrac{i}{2 U_2}\,(\vec{V}\tilde{\mathcal{A}}-\vec{\bar{V}}\mathcal{A})$. In this form, following \cite{Moore:1997pc,Stieberger:1998yi,Foerger:1998kw,Kiritsis:2000zi}, we can use modular invariance of the integrand under $\gr{0}$, and trade a modular $\Gamma_{[0]}$-transformation $\tau\mapsto\frac{a\tau+b}{c\tau+d}$ for a transformation of the matrix $A$.
This allows us to extend the domain of integration to images of $\mathbb{F}_{[0]}$ under $\gr{0}$, while simultaneously restricting the summation over $A$ to inequivalent $\gr{0}$-orbits with an appropriate choice of representative matrices. To make this more precise, we use the result of \cite{Dixon:1990pc} that a generic matrix $A$ lies in exactly one out of three inequivalent $SL(2,\mathbb{Z})$ orbits, with representatives denoted by $A=0, A_0^{\text{ND}}, A_0^{\text{D}}$ which we take to be the same as in \cite{GHPI}.

We now obtain 
\begin{align}
&\int_{\mathbb{F}_{[0]}} d^2\tau\, \tau_2  \mathcal{P}^{(k)}_0(\bar{\tau})\,
\Theta^{(2,2+k)}_{0}(\tau,\bar{\tau},\y)=\int_{\mathbb{F}_{[0]}}\frac{d^2\tau}{\tau_2^2}
\frac{Y}{U_2}\,\mathcal{P}_0^{(k)}(\bar{\tau})\sum_{\vec{\ell}\in \Gamma_{\mathfrak{g}}}\,
\bar{q}^{\frac{1}{2}\vec{\ell}\cdot \vec{\ell}}\,e^{F(0,\y)}+\nonumber\\
&+\int_{\mathbb{F}_{[0]}}\frac{d^2\tau}{\tau_2^2}
\frac{Y}{U_2}\,\mathcal{P}_0^{(k)}(\bar{\tau})\sum_{{V\in \Gamma}\atop\vec{\ell}\in \Gamma_{\mathfrak{g}}}\,
\bar{q}^{\frac{1}{2}\vec{\ell}\cdot \vec{\ell}}\,e^{F(A_0^{\text{ND}}V,\y)}+\int_{\mathbb{F}_{[0]}}\frac{d^2\tau}{\tau_2^2}
\frac{Y}{U_2}\,\mathcal{P}_0^{(k)}(\bar{\tau})\sum_{{V\in \Gamma}\atop\vec{\ell}\in \Gamma_{\mathfrak{g}}}\,
\bar{q}^{\frac{1}{2}\vec{\ell}\cdot \vec{\ell}}\,e^{F(A_0^{\text{D}}V,\y)}\,.
\end{align}
In order to apply this result to the case of $\Gamma_{[0]}\subset\Gamma$, we note that (after choosing an appropriate representative matrix $A_0$) each of these $\Gamma$-orbits can be decomposed into several (inequivalent) $\Gamma_{[0]}$ orbits by writing
\begin{align}
&A=A_0 V=\sum_{i=1}^N A_0\gamma_i\hat{V}\,,&\text{with}&&\begin{array}{l} V\in\Gamma\,, \\ \hat{V}\in\Gamma_{[0]}\,.\end{array}
\end{align}
Note moreover, that the integration over the fundamental domain of $\Gamma_{[0]}$ allows us to write 
\begin{align}
&\int_{\mathbb{F}_{[0]}} d^2\tau\, \tau_2  \mathcal{P}^{(k)}_0(\bar{\tau})\,
\Theta^{(2,2+k)}_{0}(\tau,\bar{\tau},\y)=\sum_{i=1}^N\Bigg[\int_{\mathbb{F}_{[0]}}\frac{d^2\tau}{\tau_2^2}
\frac{Y}{U_2}\,\mathcal{P}_0^{(k)}(\bar{\tau})\sum_{\vec{\ell}\in \Gamma_{\mathfrak{g}}}\,
\bar{q}^{\frac{1}{2}\vec{\ell}\cdot \vec{\ell}}\,e^{F(0,\y)}\nonumber\\
+&\int_{\mathbb{F}_{[0]}}\frac{d^2\tau}{\tau_2^2} \frac{Y}{U_2}\,\mathcal{P}_0^{(k)}(\bar{\tau})\sum_{{\hat{V}\in \Gamma_{[0]}}\atop\vec{\ell}\in \Gamma_{\mathfrak{g}}}\, \bar{q}^{\frac{1}{2}\vec{\ell}\cdot \vec{\ell}}\,e^{F(A_0^{\text{ND}}\gamma_i \hat{V},\y)}+\int_{\mathbb{F}_{[0]}}\frac{d^2\tau}{\tau_2^2}\frac{Y}{U_2}\,\mathcal{P}_0^{(k)}(\bar{\tau})\sum_{{\hat{V}\in \Gamma_{[0]}}\atop\vec{\ell}\in \Gamma_{\mathfrak{g}}}\, \bar{q}^{\frac{1}{2}\vec{\ell}\cdot \vec{\ell}}\,e^{F(A_0^{\text{D}}\gamma_i \hat{V},\y)}\Bigg]\nonumber\\
:=&\sum_{i=1}^N\left(\mathcal{I}^{[0]}_{0}(\gamma_i )+\mathcal{I}_{\text{ND}}^{[0]}(\gamma_i )+\mathcal{I}_{\text{D}}^{[0]}(\gamma_i )\right)\,.\label{IntegralSummaryResDenom}
\end{align}
For simplicity, we can focus on the trivial coset representative $\gamma_1$ which by itself results in a well defined expression. Choosing the representative matrices $A_0^{\text{ND}}$ and $A_0^{\text{D}}$ in the same way as in \cite{Dixon:1990pc} one can work out explicit expressions for the contributions to the individual orbits. 

For the zero orbit $\mathcal{I}^{[0]}_0(\gamma_1)$, 
one has $A=0$ and the integral over $\mathbb{F}_{[0]}$ can be solved using standard methods due to modular covariance with respect to $\gr{0}$, and the integral can be reduced to an integral over $\tau_1\in[-1/2,1/2)$ at $\tau_2\to\infty$. This contribution, however, is the same as the $\mu=0$ part of the zeroth orbit contribution to (\ref{TorusIntegralFinResult}).

In the non-degenerate orbit, the representative $A_0^{\text{ND}}$ can be parametrized in the standard way in terms of upper-triangular matrices with integer entries $(r, j, p)$, satisfying $r>j\geq 0$ and $p\in\mathbb{Z}\backslash \{0\}$.
%\begin{align}
%&A_0^{\text{ND}}=\left(\begin{array}{cc}r & j \\ 0 & p\end{array}\right)\,, &&\text{with}&&\left\{\begin{array}{l}p\in\mathbb{Z}\neq 0\,, \\ r>j\geq 0\,.\end{array}\right. ,
%\end{align}
The integration domain can be unfolded to the full upper half plane:
\begin{align}
\mathcal{I}_{\text{ND}}^{[0]}(\gamma_1)\sim\int_{\mathbb{H}}\frac{d^2\tau}{\tau_2^2}&\frac{Y}{U_2}\mathcal{P}_0^{(k)}\sum_{{{p\neq 0\atop r>j\geq 0}}\atop\vec{b}\in\Gamma_{\mathfrak{g}}}\,\bar{q}^{\frac{1}{2}\vec{b}\cdot \vec{b}}\,e^{F(A_0^{\text{ND}},\y)}.\nonumber
\end{align}
Up to a factor of $1/N$ this yields the $\mu=0$ contribution of non-degenerate part of (\ref{TorusIntegralFinResult}).

Finally we turn to the degenerate orbit. Realizing that transformations of the form $T^m$ for integer $m$ leave $A_0$ invariant provided we choose $c=0$ and $d=1$, we can further restrict the domain of integration to the semi-infinite strip $\mathbb{S}$ parametrised by $(\tau_1,\tau_2)\in[-1/2,1/2)\times [0,\infty)$, such that we may write
\begin{align}
\mathcal{I}_{\text{D}}^{[0]}(\gamma_1)=\int_{\mathbb{S}}\frac{d^2\tau}{\tau_2^2}&\frac{Y}{U_2}\mathcal{P}_0^{(k)}\sum_{{{j,p\in\mathbb{Z}\atop (j,p)\neq (0,0)}}\atop\vec{b}\in\Gamma_{\mathfrak{g}}}\,\bar{q}^{\frac{1}{2}\vec{b}\cdot \vec{b}}\,e^{F(A_0^{\text{D}},\y)}\,.\nonumber
\end{align}
Up to a factor of $1/N$, this is again the $\mu=0$ contribution of the degenerate part of (\ref{TorusIntegralFinResult}).

To conclude, we have found that (up to an irrelevant factor of $1/N$ which stems from the fact that there are exactly $N$ inequivalent choices of the coset representatives $\gamma_{1},\ldots,\gamma_N$) the zero conjugacy class contribution to (\ref{TorusIntegralFinResult}) can be expressed as follows
\be
\mathcal{F}_{1}^{\text{analy}}(\y)\big|_{\mu=0} \sim \mathcal{I}_0^{[0]}(\gamma_1)+\mathcal{I}_{\text{D}}^{[0]}(\gamma_1)+\mathcal{I}_{\text{ND}}^{[0]}(\gamma_1).
\ee
Hence, up to an overall $N$-dependent factor, we can write the infinite product $\Phi_{\mathfrak{g}}(\y)$ in terms of an explicit integral theta lift:
\begin{align}
\log||\Phi_{\mathfrak{g}}(\y)||^2  \sim\int_{\mathbb{F}_{[0]}} d^2\tau \, \tau_2 \mathcal{P}^{(k)}_0(\bar{\tau})\,
\Theta^{(2,2+k)}_{0}(\tau,\bar{\tau},\y)\,  +\, \cdots 
\end{align}
where the ellipsis represent irrelevant terms, c.f. (\ref{LogTermInteg}).\footnote{As a side-remark we would like to comment that the choice of the representative $\gamma_1$ was merely due to convenience. Although we have not checked this explicitly, we expect that the contributions from each of the remaining terms in (\ref{IntegralSummaryResDenom}) in fact yield similar results.} This expression makes the modular properties of $\Phi_{\mathfrak{g}}(\y)$ with respect to $\grg  \subset SO(2,2+k)$ manifest. Indeed, $\grg  $ is generically only a subgroup of the full T-duality group $SO(2,2+k;\mathbb{Z})$. More precisely, $\Phi_{\mathfrak{g}}(\y)$ retains the invariance under lattice shifts $\y \rightarrow \y +v, \, v\in \Lambda_{\mathfrak{g}^{++}}$, and under $w\in SO(1,1+k;\mathbb{Z})$, while the symmetry under $\mathcal{S}\in SO(2,2+k;\mathbb{Z})$ (see Appendix \ref{App:BorcherdsExtension}) is generically broken. These statements are consistent with Theorem 2.23 in \cite{Bruinier} (which in turn builds upon earlier work by Borcherds \cite{Borcherds1,Borcherds2}). Borcherds projects arising from lifts of Jacobi forms for $\Gamma_0(N)$ have also been constructed recently in \cite{GritsenkoClery}; it would be interesting to understand if there is a relation to our work.\footnote{We thank Boris Pioline for pointing out this reference.} 
%%%%%%%%%%%%%%%%%%%%%%%%%%%%%%%%%%%%%%%%%%%%%%%%%%%%%%%%%%%%
\subsubsection{Denominator Formula and Automorphic Correction}
With the modular properties under $\grg\subset SO(2,2+k)$ now manifest, we can indeed identify $\Phi_{\mathfrak{g}}(\y)$ with the denominator formula of a new algebra $\bor(\mathfrak{g}^{++})$. Thus, we can reinterpret the infinite product over $\Lambda_{\mathfrak{g}^{++}}^+$ in (\ref{DenomFormPosRootwoWeyl}) as a product over the positive roots $\Delta^+_{\bor}$ of $\bor(\mathfrak{g}^{++})$
\begin{align}
\Phi_{\mathfrak{g}}(\y)=e^{-2\pi i (\rho|\y)}\prod_{\alpha \in\Delta^+_{\bor}}\, \left(1-e^{2\pi i (\alpha|\y)}\right)^{c_{0}(-\alpha^2/2)}\,.\label{DenomFormPosRoot}
\end{align}
Following \cite{Borcherds0,Borcherds1}, due to the appearance of simple imaginary roots, the automorphic correction $\bor(\mathfrak{g}^{++})$ indeed falls in the class of  \emph{generalised Kac-Moody algebras}. While the multiplicity of all real positive roots is given by $c_{0}(-1)=1$, the multiplicities of the imaginary roots are encoded via (\ref{DenomFormPosRootwoWeyl}) in the remaining Fourier coefficients $c_0(-\alpha^2/2)=c_{0}\left(n'r-\vec{\ell}\cdot \vec{\ell}/2\right)$. The norm $||\cdot ||^2$ in (\ref{LogTermInteg}) can now also be interpreted as splitting the infinite product (\ref{DenomFormPosRootwoWeyl}) into contributions from positive and negative roots. More abstractly and in view of its modular properties, $\Phi_{\mathfrak{g}}(\y)$ can be interpreted as a meromorphic section of the line bundle $\mathcal{L}\rightarrow \grg   \backslash \mathcal{M}_{2,2+k}$ of weight $c_0(0)/2$ modular forms on $\mathcal{M}_{2,2+k}$. In this language, the norm $||\cdot ||$ corresponds to the invariant Petersson metric on $\mathcal{L}$ \cite{Borcherds1,Borcherds2,Kontsevich}.
%%%%%%%%%%%%%%%%%%%%%%%%%%%%%%%%%%%%%%%%%%

\section{Explicit Examples}\label{Sect:Examples}
\setcounter{equation}{0}
We shall now illustrate the general discussion of the previous
sections by a few
explicit examples for different choices of simple Lie algebras $\mathfrak{g}$. Our prime example will be the case $\mathfrak{g}=\mathfrak{a}_1$, which we will discuss in quite some detail. We will then proceed to study a list of further examples to show that our approach works in great generality. For many of these examples the decomposition of the full
$\Gamma^{2,18}$ Siegel-Narain theta-function for non-trivial $\vec{V}$ has
already been considered previously in the literature using the so called
`sequential Higgs mechanism'  \cite{Aldazabal:1995yw,Weiss:2007tk}. Our method,
however, is more flexible and allows a quick adaptation also for more general
cases. One very particular class of examples corresponding to semisimple $\mathfrak{g}$, which have not previously been
discussed in the literature, will be presented in
section~\ref{App_SemisimpleExamples}.  
%%%%%%%%%%%%%%%%%%%%%%%%%%%%%%%%%%%%%%%%%%%%%%%%%%%%%%%%%%%%%%%%%%%%%
\subsection{Simple Sequence: $\mathfrak{g}=\mathfrak{a}_k$}
The first series of examples we wish to study are $\mathfrak{g}=\mathfrak{a}_k$ with $k=1,\ldots,4$. We will be fairly explicit for the case $k=1$, which acts to demonstrate the methods we have discussed in the previous sections, and will only state the relevant results in the other cases.
%%%%%%%%%%%%%%%%%%%%%%%%%%%%%%%%%%%%%%%%%%%%%%%%%%%%%%%%%%%%%%%%%%%%%%
\subsubsection*{The case $\mathfrak{g}=\mathfrak{a}_1$}\label{Sect:ExA1}
Our first example is the case $\mathfrak{g}=\mathfrak{a}_1$ such that the unbroken gauge algebra is $\mathfrak{h}=\mathfrak{e}_7\oplus \mathfrak{e}_8$. The Wilson line $\vec{V}$ is proportional to the single root of $\mathfrak{a}_1$ and we will call the coefficient $V$ in the following. Following the work of \cite{Gannon:1991vg,Gannon:1991vj} on theta series of Lie-algebra lattices, we can immediately extract the relevant part of the integrand in (\ref{ReductionSiegelNarain}):
\begin{align}
\mathcal{P}^{(\mathfrak{a}_1)}_0(\bar{\tau})\Theta_0^{(\mathfrak{a}_1)}(\tau,\bar{\tau},V)&=\frac{E_4(\bar{\tau})^2}{\eta(\bar{\tau})^{24}}\Big(\vartheta_3(2\bar{\tau})^7+7\vartheta_3(2\bar{\tau})^3 \vartheta_2(2\bar{\tau})^4\Big)\vartheta_3(2V,2\bar{\tau}):= \frac{E_4(\bar{\tau})^2}{\eta(\bar{\tau})^{24}} f(V, \bar{\tau})\nonumber\\
&=\sum_{n=-1}^{\infty}\sum_{\ell\in \mathbb{Z}} c_0(n,\ell) \bar{q}^n\,e^{2\pi i\ell V}\,,\label{IntegrandA1}
\end{align}
where $f(V, \bar{\tau})$Ê was defined by the last equality on the first line. Explicit evaluation yields the following values for the first few Fourier coefficients
\begin{align}
&c_{0}(-1)=1 &&c_{0}(0)=630 &&c_{0}(1)=138024 &&c_{0}(2)=9987360\,.\label{coefsA1}
\end{align}
If we want to use (\ref{IntegrandA1}) to define a Borcherds extension of $\mathfrak{a}_1^{++}$ we first need to show that it transforms well under (a congruence subgroup of) $SL(2,\mathbb{Z})$. We indeed claim that (\ref{IntegrandA1}) transforms as a weak Jacobi form of weight $0$ under $\Gamma_{0}(4)$ (see appendix~\ref{App:ModFormsG04}). To see this, we first notice that the overall factor $\bar{E}_4^2/\bar{\eta}^{24}$ transforms with weight $-4$ under the full $SL(2,\mathbb{Z})$. Thus, all we have to consider are the modular properties of the function $f(V, \bar{\tau})$ defined in (\ref{IntegrandA1}). We can make the latter manifest by expanding the combination of theta series in a basis of modular forms of $\Gamma_0(4)$ (for details of the notation see appendix~\ref{App:ModFormsG04})
\begin{align}
f(V, \bar{\tau})&=\left(\frac{2E_4(\bar{\tau})}{45}-\frac{E_4(2\bar{\tau})}{20}+\frac{4E_4(4\bar{\tau})}{45}\right)\phi_{0,1}(\bar{\tau},V)\nonumber\\
&-\left(\frac{8E_6(\bar{\tau})}{189}-\frac{11E_6(2\bar{\tau})}{252}+\frac{16E_6(4\bar{\tau})}{189}+26h_6(\bar{\tau})\right)\phi_{-2,1}(\bar{\tau},V)\,,
\end{align}
which indeed proves that (\ref{IntegrandA1}) is invariant under $\Gamma_0(4)$. This implies that the infinite product $\Phi_{\mathfrak{a}_1}(\y)$, which is computed from (\ref{Phidef}) by inserting the coefficients (\ref{coefsA1}), is also automorphic with respect to a finite index subgroup $\grg  \subset SO(2,3;\mathbb{Z})$ which is induced by $\Gamma_0(4)$ through the theta correspondence \cite{Bruinier} (see also section 13 of \cite{Borcherds1} for a discussion of modular products induced from modular forms for $\Gamma_0(N)$). Therefore, as explained before, $\Phi_{\mathfrak{a}_1}(\y)$ defines the denominator formula for a BKM algebra $\bor({\mathfrak a}_1^{++})$, {\it i.e.} 
\begin{align}
\Phi_{\mathfrak{a}_1}(\y)=e^{-2\pi i (\rho|\y)}\prod_{\alpha \in\Delta^{+}_{\bor({\mathfrak a}_1^{++})}}\, \left(1-e^{2\pi i (\alpha|\y)}\right)^{c_{0}\left(-(\alpha|\alpha)/2\right)}\, .\label{DenomFormPosRootwoWeylA1}
\end{align}
The root multiplicities of $\bor({\mathfrak a}_1^{++})$ are simply given by ${\rm mult} (\alpha) = c_{0}\left( - \alpha^2/2  \right) 
= c_{0}( n' r - \tfrac{1}{2} \vec{\ell} \cdot \vec{\ell} \,)$, encoded in (\ref{coefsA1}). In particular, as we can read off, the simple positive roots all have squared length $2$, and thus appear with multiplicity $c_{0}(-1)=1$. The corresponding hyperbolic subalgebra $\mathfrak{a}_{k}^{++}$ is characterised by the $3\times3$ Cartan matrix whose Dynkin diagram
is: 
\begin{align}
{\begin{picture}(0,0)%
\includegraphics{Dyna1pp.pstex}%
\end{picture}%
\setlength{\unitlength}{4144sp}%
\begingroup\makeatletter\ifx\SetFigFontNFSS\undefined%
\gdef\SetFigFontNFSS#1#2#3#4#5{%
  \reset@font\fontsize{#1}{#2pt}%
  \fontfamily{#3}\fontseries{#4}\fontshape{#5}%
  \selectfont}%
\fi\endgroup%
\begin{picture}(1015,283)(2528,-3748)
\put(2543,-3698){\makebox(0,0)[lb]{\smash{{\SetFigFontNFSS{6}{7.2}{\rmdefault}{\mddefault}{\updefault}{\color[rgb]{0,0,0}$\alpha_{-1}$}%
}}}}
\put(2981,-3702){\makebox(0,0)[lb]{\smash{{\SetFigFontNFSS{6}{7.2}{\rmdefault}{\mddefault}{\updefault}{\color[rgb]{0,0,0}$\alpha_{0}$}%
}}}}
\put(3404,-3698){\makebox(0,0)[lb]{\smash{{\SetFigFontNFSS{6}{7.2}{\rmdefault}{\mddefault}{\updefault}{\color[rgb]{0,0,0}$\alpha_{1}$}%
}}}}
\end{picture}%
}\, 
\end{align}
As a final comment, we would like to remark that $\bor(\mathfrak{a}_1^{++})$ constructed here differs from the automorphic completion $\mathfrak{g}_{1,0}$ of $\mathfrak{a}_1^{++}$ considered in \cite{Gritsenko:1996ax} since we have not added any odd roots; in other words, $\bor(\mathfrak{a}_1^{++})$ is not a `super BKM-algebra', in contrast to $\mathfrak{g}_{1,0}$.
%%%%%%%%%%%%%%%%%%%%%%%%%%%%%%%%%%%%%%%%%%%%%%%%%%%%%%%%%%%%%%%%%%%%%%%%%%
\subsubsection*{The cases $\mathfrak{g}=\mathfrak{a}_2, \mathfrak{a}_3, \mathfrak{a}_4$}
Let us now also briefly sketch the remaining members of this series of Lie algebras, \emph{i.e.} the examples $\mathfrak{g}=\mathfrak{a}_k\cong \mathfrak{sl}(k+1,\mathbb{R})$ for $k=2,3,4$
with the algebras $\mathfrak{h}$ given by $\mathfrak{e}_6\oplus \mathfrak{e}_8$,
$\mathfrak{d}_5\oplus \mathfrak{e}_8$ 
and $\mathfrak{a}_4\oplus \mathfrak{e}_8$ respectively. Furthermore we will use the notation 
$y=(U,T;V_{i(k)})$.\footnote{In order to avoid cluttering the notation, we will
from now on denote $V_{i(k)}$ simply by $V_i$.} We are again interested in the contribution of the zero-conjugacy class in the integrand (\ref{ReductionSiegelNarain}). The latter can be derived in a straight-forward manner using the results of \cite{Gannon:1991vg,Gannon:1991vj,Conway} for the theta-series of the root lattices of $\mathfrak{a}_k$, $\mathfrak{d}_5$ and $\mathfrak{e}_6$. 

Indeed, with this information, we can immediately compute the Fourier-coefficients $c_{\mu}$, introduced in (\ref{FourierExpansionConjugacy}). For the reader's convenience, we have compiled the first few of them in table~\ref{Tab:FourierCoef}. 
\begin{table}[htb]
\begin{center}
\begin{tabular}{|c|llll|}\hline
\multicolumn{1}{|c|}{$\mathfrak{g}$ } & \multicolumn{4}{|c|}{\textbf{Fourier 
coefficients
}$c_{0}(n-\tfrac{1}{2}\,\vec{\ell}\cdot\vec{\ell})$}\\
\hline
&&&&\\[-10pt]
$\mathfrak{a}_2$ & $c_{0}(-1)=1$ & $c_{0}(0)=576$ & $c_{0}(1)=110322$ &
$c_{0}(2)=8142848$\\ 
&&&&\\[-10pt]
%& $c_{1}(-1/3)=27$ & $c_{1}(2/3)=13824$ & $c_{1}(5/3)=2100951$ &
%$c_{1}(8/3)=88931520$\\
%&&&&\\[-10pt]
\hline
&&&&\\[-10pt]
$\mathfrak{a}_3$ & $c_{0}(-1)=1$ & $c_{0}(0)=544$ & $c_{0}(1)=94014$ &
$c_{0}(2)=5691200$\\
&&&&\\[-10pt]
%& $c_{1}(-1/2)=10$ & $c_{1}(1/2)=5120$ & $c_{1}(3/2)=778072$ &
%$c_{1}(5/2)=32908288$\\
%&&&&\\[-10pt]
%& $c_{2}(-3/8)=16$ & $c_{2}(5/8)=8144$ & $c_{2}(13/8)=1220704$ &
%$c_{2}(21/8)=49102640$\\
%&&&&\\[-10pt]
\hline
&&&&\\[-10pt]
$\mathfrak{a}_4$ & $c_{0}(-1)=1$ & $c_{0}(0)=524$ & $c_{0}(1)=83874$ &
$c_{0}(2)=4185500$ \\[10pt] \hline
%& $c_{1}(-3/5)=5$ & $c_{1}(2/5)=2550$ & $c_{1}(7/5)=383970$ &
%$c_{1}(12/5)=15703320$\\
%&&&&\\[-10pt]
%& $c_{2}(-2/5)=10$ & $c_{2}(3/5)=5065$ & $c_{2}(8/5)=750300$ &
%$c_{2}(13/5)=28824775$\\[10pt] 
\end{tabular}
\caption{Fourier coefficients of (\ref{Jacobi}) for various algebras $\mathfrak{g}$ which can be deduced using the work of \cite{Gannon:1991vg,Gannon:1991vj}.}
\label{Tab:FourierCoef}
\end{center}
\end{table}
The explicit expressions of the theta-series also imply modular invariance of the $\mu=0$ contribution under some particular congruence subgroup of $SL(2,\mathbb{Z})$, which is again necessary for an interpretation as the denominator formula for a BKM algebra. To be specific, the groups are given in the following table:
\begin{center}
\begin{tabular}{|c|cccc|}\hline
\textbf{group} & $\mathfrak{a}_1$&$\mathfrak{a}_2$ & $\mathfrak{a}_3$ & $\mathfrak{a}_4$ \\ \hline  & &&&\\[-10pt]
$\Gamma_{[0]}$ & $\Gamma_0(4)$ &$\Gamma_{0}(6)$ & $\Gamma_{0}(8)$ & $\Gamma_0(10)$ \\[6pt] \hline
\end{tabular}
\end{center}
As explained before, this implies that $\Phi_{\mathfrak{a}_k}(\y)$, which is defined using the coefficients of table~\ref{Tab:FourierCoef} has good modular properties under a finite index subgroup $\grg  \subset SO(2,2+k;\mathbb{Z})$ and can be interpreted as the denominator formula for the BKM algebra $\bor({\mathfrak a}_k^{++})$
\begin{align}
\Phi_{\mathfrak{a}_k}(\y)=e^{-2\pi i (\rho|\y)}\prod_{\alpha \in\Delta^{+}_{\bor({\mathfrak a}_k^{++})}}\,
 \left(1-e^{2\pi i (\alpha|\y)}\right)^{c_{0}\left(-(\alpha|\alpha)/2\right)}\, .\label{DenomFormPosRootwoWeyl1}
\end{align}
Thus the root multiplicities of $\bor({\mathfrak a}_k^{++})$ are simply given by the Fourier coefficients in Table \ref{Tab:FourierCoef}. In particular, the simple positive roots all have length $2$, and thus appear
with multiplicity $c_{0}(-1)=1$. The corresponding hyperbolic subalgebra $\mathfrak{a}_{k}^{++}$ is characterised
by the 
$(k+2)\times (k+2)$ Cartan matrix 
whose Dynkin diagram
is of the form
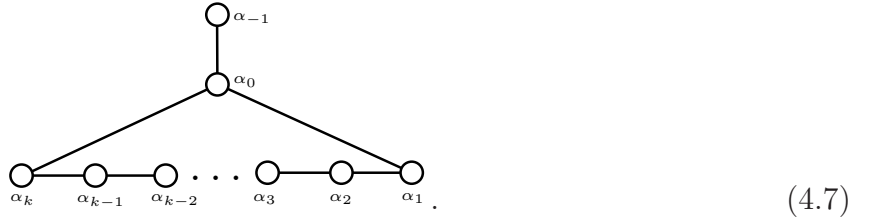
\begin{align}
{\begin{picture}(0,0)%
\includegraphics{Dynakpp.pstex}%
\end{picture}%
\setlength{\unitlength}{4144sp}%
\begingroup\makeatletter\ifx\SetFigFontNFSS\undefined%
\gdef\SetFigFontNFSS#1#2#3#4#5{%
  \reset@font\fontsize{#1}{#2pt}%
  \fontfamily{#3}\fontseries{#4}\fontshape{#5}%
  \selectfont}%
\fi\endgroup%
\begin{picture}(2471,1243)(2528,-3748)
\put(2543,-3698){\makebox(0,0)[lb]{\smash{{\SetFigFontNFSS{6}{7.2}{\rmdefault}{\mddefault}{\updefault}{\color[rgb]{0,0,0}$\alpha_k$}%
}}}}
\put(3374,-3698){\makebox(0,0)[lb]{\smash{{\SetFigFontNFSS{6}{7.2}{\rmdefault}{\mddefault}{\updefault}{\color[rgb]{0,0,0}$\alpha_{k-2}$}%
}}}}
\put(3601,-3571){\makebox(0,0)[lb]{\smash{{\SetFigFontNFSS{6}{7.2}{\rmdefault}{\mddefault}{\updefault}{\color[rgb]{0,0,0}{\Large$\ldots$}}%
}}}}
\put(2937,-3702){\makebox(0,0)[lb]{\smash{{\SetFigFontNFSS{6}{7.2}{\rmdefault}{\mddefault}{\updefault}{\color[rgb]{0,0,0}$\alpha_{k-1}$}%
}}}}
\put(3980,-3694){\makebox(0,0)[lb]{\smash{{\SetFigFontNFSS{6}{7.2}{\rmdefault}{\mddefault}{\updefault}{\color[rgb]{0,0,0}$\alpha_{3}$}%
}}}}
\put(4859,-3691){\makebox(0,0)[lb]{\smash{{\SetFigFontNFSS{6}{7.2}{\rmdefault}{\mddefault}{\updefault}{\color[rgb]{0,0,0}$\alpha_{1}$}%
}}}}
\put(4432,-3694){\makebox(0,0)[lb]{\smash{{\SetFigFontNFSS{6}{7.2}{\rmdefault}{\mddefault}{\updefault}{\color[rgb]{0,0,0}$\alpha_{2}$}%
}}}}
\put(3865,-2994){\makebox(0,0)[lb]{\smash{{\SetFigFontNFSS{6}{7.2}{\rmdefault}{\mddefault}{\updefault}{\color[rgb]{0,0,0}$\alpha_{0}$}%
}}}}
\put(3862,-2615){\makebox(0,0)[lb]{\smash{{\SetFigFontNFSS{6}{7.2}{\rmdefault}{\mddefault}{\updefault}{\color[rgb]{0,0,0}$\alpha_{-1}$}%
}}}}
\end{picture}%
}\,.
\end{align}
\subsection{Semisimple Sequence: $\mathfrak{g}=\mathfrak{a}_{k_1}\oplus\mathfrak{a}_{k_2}$}\label{App_SemisimpleExamples}
\begin{figure}[ht]
\begin{center}
\input{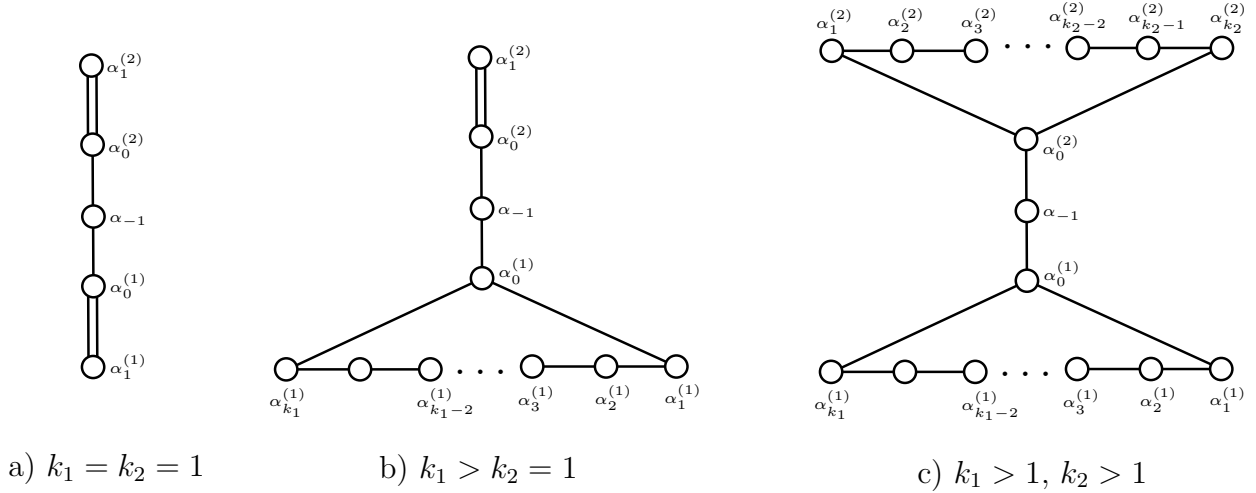}
\caption{Dynkin diagrams of the double extensions 
$(\mathfrak{a}_{k_1}\oplus \mathfrak{a}_{k_2})^{++}$ for various $(k_1,k_2)$.}
\label{DynkinAkAkpp}
\end{center}
\end{figure}

We also want to discuss examples in which $\mathfrak{g}$ is no longer a simple group. As an illustrative series of examples let us consider the case where 
\begin{align}
\begin{array}{l}\mathfrak{g}=\mathfrak{a}_{k_1}\oplus\mathfrak{a}_{k_2} \\
\mathfrak{h}=\mathfrak{h}_{k_1}\oplus\mathfrak{h}_{k_2} \end{array}&&
\text{with}&&\mathfrak{h}_{k_i}=\left\{\begin{array}{lcl}\mathfrak{e}_7 & \text{if} &
k_i=1 \\ \mathfrak{e}_6 & \text{if} & k_i=2 \\ \mathfrak{d}_5 & \text{if} &
k_i=3 \\ \mathfrak{a}_4 & \text{if} & k_i=4 \end{array}\right. && (i=1,2) \ .
\end{align}
 In order to be able to make use of the results of
section~\ref{Sect:TorusIntegralEval} we need to find the equivalent of the Fourier coefficients
introduced in (\ref{FourierExpansionConjugacy}). To this end we perform a Poisson-resummation, after which we can write for the integral the following sum over conjugacy classes
\begin{align}
\mathcal{F}_{1}^{\text{analy}}(\y)=\int_{\mathbb{F}}\frac{d^2\tau}{\tau_2^2}&\frac{Y}{U_2}\sum_{\mu=0}^{s-1}\frac{G_2^{\text{analy}}}{\bar{\eta}^{24}}\,\sum_{\mu=0}^{s-1}
\Theta^{\mathfrak{h}_{k_1}}_{\mu}(\bar{\tau})\, 
\Theta^{\mathfrak{h}_{k_2}}_{\mu}(\bar{\tau})\,
\sum_{{(p_1,n_1;p_2,n_2)}}\sum_{{\vec{\ell}_1\in\Lambda_{1}+\lambda_\mu^1}\atop
\vec{\ell}_2\in\Lambda_{2}+\lambda_\mu^2}\,\bar{q}^{\frac{1}{2}\left(\vec{\ell}_1\cdot
\vec{\ell}_1+\vec{\ell}_2\cdot \vec{\ell}_2\right)}\cdot\nonumber\\
&\times e^{2\pi i\vec{\ell}_1\cdot \vec{z}_1+2\pi i\vec{\ell}_2\cdot
\vec{z}_2-\frac{\pi Y}{U_2^2\tau_2}|\mathcal{A}|^2-2\pi iT\text{det}A-\frac{\pi
n_2\left(\vec{V}^2\tilde{\mathcal{A}}-\vec{\bar{V}}^2\mathcal{A}\right)}{U_2}
+\frac{2\pi i\,(\Im\vec{V})^2}{U_2^2}(n_1+n_2\bar{U})\mathcal{A}}\,,
\end{align}
Here $\lambda_\mu^1$ and $\lambda_\mu^2$ are the projections of the glue vector
on the root lattices $\Lambda_{1}$ and $\Lambda_{2}$ of $\mathfrak{a}_{k_1}$ and $\mathfrak{a}_{k_2}$, 
respectively, while $\Theta^{\mathfrak{h}_{k_{1,2}}}_{\mu}(\bar{\tau})$ are the theta-series of the various $\Lambda_{\mathfrak{h}_{k_{1,2}}}$ cosets. The latter are obtained from the projections $\lambda^{\mathfrak{h}_{1,2}}_\mu$ of the glue vector $\lambda_\mu$ onto $\mathfrak{h}_{k_1}$ and $\mathfrak{h}_{k_2}$ respectively
\begin{align}
&\Theta^{\mathfrak{h}_{k_{a}}}_{\mu}(\bar{\tau})=\sum_{\vec{\ell}\in\Lambda_{\mathfrak{h}_{k_a}}+\lambda_\mu^{\mathfrak{h}_{k_a}}}\bar{q}^{\frac{1}{2}\,\vec{\ell}\cdot\vec{\ell}}\,,&&\forall a=1,2\,.
\end{align}
As before, $\mu=0,\ldots,s-1$ labels the various conjugacy classes.
The Fourier expansion (\ref{FourierExpansionConjugacy}) for the case at hand can
then be written more explicitly as
\begin{align}
\frac{G_2^{\text{analy}}}{\bar{\eta}^{24}}&\,\sum_{\mu=0}^{s-1}
\Theta^{\mathfrak{h}_{k_1}}_{\mu}(\bar{\tau})\, 
\Theta^{\mathfrak{h}_{k_2}}_{\mu}(\bar{\tau})\,
\sum_{{(p_1,n_1;p_2,n_2)}}\sum_{{\vec{\ell}_1\in\Lambda_{1}+\lambda_\mu^{1}}\atop
\vec{\ell}_2\in\Lambda_{2}+\lambda_\mu^{2}}\,\bar{q}^{\frac{1}{2}\left(\vec{\ell}_1\cdot
\vec{\ell}_1+\vec{\ell}_2\cdot \vec{\ell}_2\right)}\, e^{2\pi i\vec{\ell}_1\cdot
\vec{z}_1+2\pi i\vec{\ell}_2\cdot \vec{z}_2}=\nonumber\\
&=\sum_{\mu=0}^{s-1}\sum_{n=-1}^\infty\sum_{{\vec{\ell}_1\in\Lambda_{1}
+\lambda_\mu^1}\atop\vec{\ell}_2\in\Lambda_{2}+\lambda_\mu^2}
c_{\mu}\left[n-\tfrac{1}{2}\left(\vec{\ell}_1\cdot\vec{\ell}_1+\vec{\ell}_2\cdot\vec{\ell}_2\right)\right]\,
\bar{q}^ne^{2\pi i\left(\vec{\ell}_1\cdot\vec{z}_1+\vec{\ell}_2\cdot\vec{z}_2\right)} \ .
\end{align}
The modular properties particularly of the $\mu=0$ contribution follow this time already from our analysis of section~\ref{Sect:ExA1} and we can thus immediately proceed to interpretation in terms of a BKM algebra. According to section~\ref{Sect:DenominatorFormula}, the important information
about root-multiplicities ({\it i.e.}\ the denominator formula) of
$\bor(\mathfrak{g}^{++})$ is encoded in the Fourier-coefficients of the trivial
conjugacy class $\mu=0$. We have tabulated the first few such coefficients for different choices
of $k_1$ and $k_2$ in Table~\ref{Tab:FourierCoefSemi}.
\begin{table}[htb]
\begin{center}
\begin{tabular}{|c|r|l|l|l|l|}\hline
& $n$ & $\mathfrak{a}_1$ & $\mathfrak{a}_2$ & $\mathfrak{a}_3$ &
$\mathfrak{a}_4$\\\hline
$\mathfrak{a}_1$ & $-1$ & $1$ & $1$ & $1$ & $1$\\
& $0$ & $516$ & $462$ & $430$ & $410$\\
& $1$ & $92160$ & $70614$ & $57954$ & $50094$\\
& $2$ & $7002096$ & $4528948$ & $3105820$ & $2236960$\\\hline
$\mathfrak{a}_2$ & $-1$ & & $1$ & $1$ & $1$\\
& $0$ & & $408$ & $376$ & $356$\\
& $1$ & & $51984$ & $41052$ & $34272$\\
& $2$ & & $2878112$ & $1936448$ & $1365668$\\\hline
$\mathfrak{a}_3$ & $-1$ &  &  & $1$ & $1$\\
& $0$ & & & $344$ & $324$\\
& $1$ & & & $31144$ & $25004$\\
& $2$ & & & $1276640$ & $880340$\\\hline
$\mathfrak{a}_3$ & $-1$ &  &  &  & $1$\\
& $0$ & & & & $304$\\
& $1$ & & & & $19264$\\
& $2$ & & & & $592040$\\\hline
\end{tabular}
\caption{Fourier coefficients $c_{0}(n)$ for various algebras
$\mathfrak{g}=\mathfrak{g}_{(1)}\oplus\mathfrak{g}_{(2)}$. Note that the coefficients are symmetric under the exchange of $\mathfrak{g}_{(1)}$ and $\mathfrak{g}_{(2)}$.}
\label{Tab:FourierCoefSemi}
\end{center}
\end{table}
These coefficients are
sufficient to obtain the full denominator formula from (\ref{LogTermInteg}).
Let us consider this result also from a more algebraic perspective,
{\it i.e.}\ from the point of view of equation (\ref{DenomFormPosRootwoWeyl}). The 
$(2+k_1+k_2)\times (2+k_1+k_2)$ Cartan matrix
is encoded in the Dynkin diagrams in figure~\ref{DynkinAkAkpp}.
This matrix
has a single zero eigenvalue, which implies that the associated
simple roots are not all linearly independent. Indeed one can check that
\begin{align}
\alpha_0^{(2)}=\alpha_0^{(1)}-\sum_{i=1}^{k_1}\alpha_i^{(1)}+\sum_{m=1}^{k_2}
\alpha_m^{(2)}\ .
\end{align}
Following the
general discussion of section~\ref{Sect:DenominatorFormula}, the Fourier
coefficients $c_{0}$ will correspond to the root multiplicities of the
algebra\footnote{Our conventions for double extensions of semisimple Lie algebras follow the philosophy of \cite{Kleinschmidt:2008jj}; see also Appendix A of \cite{GHPI} for details on our precise conventions.} $\bor((\mathfrak{a}_{k_1}\oplus\mathfrak{a}_{k_2})^{++})$. As required the simple positive roots all have length $2$ and therefore have multiplicity
$c_{0}(-1)=1$.\\

%%%%%%%%%%%%%%%%%%%%%%%%%%%%%%%%%%%%%%%%%%%%%%%%%%%%%%%%%%%%%%%%%%%%%%%%%%%%%%%%
%%%%%%%%%%%%%%%%%%%%%%%%%

%%%%%%%%%%%%%%%%%%%%%%%%%%%%%%%%%%%%%%%%%%%%%%%%%%%%%%%%%%%%%%%%%%%%%%%%%%%%%
%%%%%%%%%%%%%%%%%%%%%%%%%%%%%%%%%%%%%%%%%%%%%%%%%%%%%%%%%%%%%%%%%%%%%%%%%%%%%

%%%%%%%%%%%%%%%%%%%%%%%%%%%%%%%%%%%%%%%%%%%%%%%%%%%%%%%%%%%%%%%%%%%%%%%%%%%%%%%%
%%%%%%%%%%%%%%
\section{Conclusions and Discussion}\label{Conclusions}

In this work we have analysed a particular $\cN=4$ topological one-loop amplitude $\mathcal{F}_1$
in heterotic string theory on $\mathbb{T}^2$. We evaluated the integral $\mathcal{F}_1$ explicitly for arbitrary enhanced semisimple gauge group $ \mathfrak{h}\subset \mathfrak{e}_8\oplus \mathfrak{e}_8$, i.e. for any choice of Wilson lines. The analytic part $\mathcal{F}_{1}^{\text{analy}}(\y)$ can be written in terms of an infinite product over a Lorentzian lattice, identified with the root lattice of the Lorentzian extension 
$\mathfrak{g}^{++}$ of the complement $\mathfrak{g}= (\mathfrak{e}_8\oplus\mathfrak{e}_8)/\mathfrak{h}$. Using the Borcherds-Gritsenko-Nikulin philosophy of ``automorphic correction'', this gives rise to a class of 
Borcherds algebras $\mathcal{G}(\mathfrak{g}^{++})$, of which the root multiplicities
are explicitly calculable in terms of the Fourier coefficients of certain modular forms. 

%In the particular example where precisely one of the $\mathfrak{e}_8$-factors is
%completely broken, one has
%$\mathcal{G}({\mathfrak{g}}^{++})=\mathcal{G}(\mathfrak{e}_{10})$, namely a Borcherds
%extension of the hyperbolic Kac-Moody algebra $\mathfrak{e}_{10}$. This recovers
%a previous result of Harvey and Moore in the $\mathcal{N}=2$ context \cite{Harvey:1995fq}. 

As a by-product of our analysis we have provided explicit expressions for this
class of one-loop integrals in heterotic string theory on $\mathbb{T}^2$ for an
arbitrary breaking of the gauge group. In particular, our method does not
require the factors $\mathfrak{h}$ and $\mathfrak{g}$ in (\ref{DecomposeLie}) 
to be simple. These results 
generalise previous work which had been restricted to specific choices of
Wilson lines, notably always keeping one of the $\mathfrak{e}_8$-factors
unbroken (see for instance \cite{Harvey:1995fq,LopesCardoso:1996nc,Weiss:2007tk}).

The present work arose as a continuation of our previous analysis \cite{GHPI}, where 
a certain universal 'algebra of BPS states' $\mathcal{G}$ for heterotic string theory on $\mathbb{T}^2$ 
was constructed using an auxiliary bosonic conformal field theory. It is an interesting open question 
whether there is a similar 'microscopic' CFT construction of the class of 
'automorphically corrected' Borcherds algebras $\mathcal{G}(\mathfrak{g}^{++})$  uncovered herein. Although these algebras appear not 
to be subalgebras of the BPS-algebra of \cite{GHPI}, it is conceivable that they can be obtained as quotients of $\mathcal{G}$.

A natural extension of our analysis would be to go away from the large volume limit of $\mathbb{T}^4$ and consider the full $\mathcal{N}=4$ amplitude on $\mathbb{T}^6$ for which the Narain moduli space is enlarged to $SO(6, 22;\mathbb{Z})\backslash \mathcal{M}_{6,22}$. Since this is no longer a Hermitian symmetric domain, one might recover a complex structure by treating the harmonic superspace amplitude $\mathcal{F}_g(\y, u, \bar{u})$ as an automorphic function on the extended moduli space $\mathcal{M}_{6,22}\times SU(4)/(SU(2)\times SU(2)\times U(1))\cong SO(6,22)/(SO(4)\times SO(2)\times SO(22))$, similarly to the twistor space construction of \cite{Michel:2008bx}.\footnote{We thank Boris Pioline for suggesting this possibility.} This point of view might also shed light on the geometric meaning of the harmonicity and second order equations satisfied by $\mathcal{F}_g(\y, u, \bar{u})$.

%%%%%%%%%%%%%%%%%%%%%%%%%%%%%%%%%%%%%%%%%%%%%%%%%%%%%%%%%%%%%%%%%%%%%%%%%%
\vspace{.5cm}

\noindent{\bf \large{Acknowledgments}}

\vspace{.1cm}

\noindent We are especially grateful to Matthias Gaberdiel for collaboration in the initial stages of this work, as well as for numerous invaluable discussions and useful comments on an earlier draft. We are also indebted to Boris Pioline for many helpful discussions and for providing comments on a previous draft. Finally, we thank Jeff Harvey, Marcos Mari$\tilde{\text{n}}$o, Greg Moore and Roberto Volpato for stimulating discussions and correspondence. SH would like to thank ETH Z\"urich for kind hospitality during the final stages of this work. This work was partially supported by the Swiss National Science Foundation.

\appendix

\section{Weyl Vectors and Denominator Formulas}\label{App:BorcherdsExtension}
\renewcommand{\theequation}{\Alph{section}.\arabic{equation}}
\setcounter{equation}{0}
Our conventions for Lie algebras and Borcherds-Kac-Moody algebras can be found in Appendix A of \cite{GHPI}. Below we just briefly recall some of the essential features which are needed for the present analysis. We denote by $\mathfrak{g}$ a finite Lie algebra, $\mathfrak{g}^{++}$ its Lorentzian extension, and by $\bor$ a general BKM-algebra.

Similarly as for finite Lie algebras, BKM-algebras have a Weyl
vector $\rho$, satisfying $(\rho|\alpha)\leq -\frac{1}{2}(\alpha|\alpha)$,
with equality if and only if $\alpha$ is a simple root. By restricting the general Weyl-Kac-Borcherds character formula to the trivial representation one obtains the so called \emph{denominator formula}
\be
\sum_{w\in \mathcal{W}} \epsilon(w)w(S)=e^{\rho}
\prod_{\alpha\in \Delta_+}\left(1-e^{\alpha}\right)^{\text{mult}\, \alpha}\ .
\label{denominatorformula}
\ee
This formula relates a sum over the Weyl group $\mathcal{W}(\mathcal{G})$ to an
infinite product over all positive roots $\Delta_+$ of $\mathcal{G}$. The factor $S(w)$ is a
correction due to the imaginary simple roots \cite{Borcherds0}:
\be
S=e^{\lambda+\rho}\sum_{\alpha\in \Lambda_\mathcal{G}^{+}} \xi(\alpha)e^{\alpha}
\, ,
\ee
where $\xi(\alpha)=(-1)^{m}$ if $\alpha$ is a sum of $m$ distinct pairwise
orthogonal 
imaginary simple roots which are orthogonal to $\lambda$, and $\xi(\alpha)=0$
otherwise.

A key point of this paper is the fact that
BKM algebras 
$\bor$ can be constructed from Lorentzian Kac-Moody algebras
$\mathfrak{g}^{++}$ through a so-called \emph{automorphic correction} \cite{Borcherds1,GritsenkoNikulin}, as we now recall.

\noindent Suppose we are given a weak Jacobi form $\psi(\tau,z)$ with expansion
coefficients 
\begin{align}
\psi_{\mathfrak{g}}(\tau,z)=\sum_{\lambda\in\Lambda_{\mathfrak{g}^{++}}}c(\lambda)\,
q^{-\tfrac{1}{2}(\lambda|\lambda)} e^{2\pi i (z|\lambda)} \ ,\label{fExpand}
\end{align}
where $\Lambda_{\mathfrak{g}^{++}}$ is the root lattice of  
$\mathfrak{g}^{++}$. Then we consider the \emph{modular product}
\cite{Borcherds1} 
\begin{equation}
\Phi_{\mathfrak{g}}(\y)=e^{-2\pi i (\rho|\y)}\prod_{\lambda\in\Lambda_{\mathfrak{g}^{++}}^+}
(1-e^{-2\pi i(\lambda|\y)})^{c(\lambda)}\ ,\qquad 
\y\in\Lambda_{\mathfrak{g}^{++}}\otimes \mathbb{C}\ ,\label{AutoCorrect}
\end{equation}
where $\rho$ is the lattice Weyl-vector of ${\mathfrak{g}^{++}}$. 
Upon identifying (\ref{AutoCorrect}) with  (\ref{denominatorformula}) we
interpret the additional 
terms in (\ref{AutoCorrect}) with additional roots --- beyond those already in 
$\Delta^+_{\mathfrak{g}^{++}}$. Because of the crucial minus sign in the
exponent of 
$q$ in (\ref{fExpand}), these additional roots are generically
imaginary.\footnote{We do not 
consider the case in this paper that also additional simple real roots are added
in this 
way. See \cite{Govindarajan:2008vi} for examples where this happens.} It was shown in \cite{Borcherds0,Borcherds1} that there exists indeed a
BKM $\bor$ 
with these roots. 
In order to emphasize that the latter was constructed from $\mathfrak{g}^{++}$ 
we will in many cases write $\bor\equiv \bor(\mathfrak{g}^{++})$. It is,
however, important to 
realise that the extension of $\mathfrak{g}^{++}$ is not unique, since different
modular products 
will lead to different algebras $\bor$. 

By construction, the product $\Phi_{\mathfrak{g}}(\y)$ is an automorphic function for $SO(1,1+k;\mathbb{Z})$. However, 
Borcherds shows \cite{Borcherds1} that in fact $\Phi_{\mathfrak{g}}(\y)$ extends to an automorphic form
of weight $c(0)/2$ for the full T-duality group $SO(2,2+k;\mathbb{Z})$. To be precise, it is invariant under shifts 
\begin{align}
&\Phi_{\mathfrak{g}}(\y+v)=\Phi_{\mathfrak{g}}(\y)\,,&&\text{with} &&v \in \Lambda_{\mathfrak{g}^{++}}\,,
\end{align}
and arbitrary transformations under $SO(1,1+k;\mathbb{Z})$ (maybe even extended by a non-trivial multiplier system, see \emph{e.g.} \cite{Borcherds1})
\begin{align}
&\Phi_{\mathfrak{g}}\big(w(\y)\big)=\Phi_{\mathfrak{g}}(\y)\,,&&\text{with} &&w\in SO(1,1+k;\mathbb{Z})\,.
\end{align}
However, it transforms with weight $c(0)/2$ under the following transformation
\begin{align}
&\Phi_{\mathfrak{g}}\big(\mathcal{S}(\y)\big)=\left[\frac{(\y|\y)}{2}\right]^{c(0)/2}\Phi_{\mathfrak{g}}(\y)\,,&&\text{with} &&SO(2,2+k;\mathbb{Z})\ni\mathcal{S}:\,\y\mapsto  \frac{2\y}{(\y|\y)}\, .
\label{functionaleq}
\end{align}
More generally, if $\Phi_{\mathfrak{g}}(\y)$ is a modular form for a subgroup $\grg \subset SO(2,2+k;\mathbb{Z})$, the weight under the corresponding $\mathcal{S}$-transformation is given by a character of $\grg$ (see, e.g. \cite{Bruinier}).

%Let us end this subsection by computing the Weyl vector $\rho$ for the 
%BKM algebra $\mathcal{G}(\mathfrak{g}^{++})$. If ${\mathfrak g}$ is simple, one
%can show that the Weyl vector takes the form 
%\begin{align}
%\rho=(\rho_{-1},\rho_0;\vec{\rho})\,,\label{AnsatzWeylSimple}
%\end{align}
%where $\vec{\rho}$ is the Weyl vector of $\mathfrak{g}$. One can check that the condition (\ref{DefEquWeyl}) is obeyed provided 
%\begin{equation}
%\rho_{-1} = \bigl(\vec{\rho}|\vec{\theta}\bigr)-2  \qquad
%\rho_0 = \bigl(\vec{\rho}|\vec{\theta}\bigr)-1 \, , 
%\end{equation}
%where $\vec{\theta}$ is the highest root of $\mathfrak{g}$.

%When $\mathfrak{g}$ is semisimple, $\mathfrak{g}=\mathfrak{g}_{(1)}\oplus \mathfrak{g}_{(2)}$ (for simplicity we restrict to only two factors), we instead make
%the following ansatz for the Weyl vector
%\begin{align}
%&\rho=(\rho_{-1},\rho_0;\rho_1^{(1)},\ldots,\rho_{k_1}^{(1)};\rho_1^{(2)},\ldots
%,\rho_{k_2}^{(2)})\,,&&\text{with} &&\rho_I\in\mathbb{R}\,.
%\end{align}
%Evaluation of (\ref{DefEquWeyl}) for the roots
%(\ref{SetRootsOver1})-(\ref{SetRootsOver5}) yields the following explicit expression 
%\begin{align}
%\rho=\left((\vec{\rho}_1|\vec{\theta}_{(1)})-1,(\vec{\rho}_1|\vec{\theta}_{(1)}
%)-2;\vec{\rho}_{(1)};\vec{\rho}_{(2)}\right)\,,
%\end{align}
%where $(\vec{\rho}_{(1)},\vec{\rho}_{(2)})$ are the Weyl vectors of
%$(\mathfrak{g}_{(1)},\mathfrak{g}_{(2)})$.

%%%%%%%%%%%%%%%%%%%%%%%%%%%%%%%%%%%%%%%%%%%%%%%%%%%%%%%%%%%%%%%%%%%%%%%%%%%%%%%%%%%%%%
\section{Positive Root Condition}\label{Sect:PosRootCondition}
\setcounter{equation}{0}
%%%%%%%%%%%%%%%%%%%%%%%%%%%%%%%%%%%%%%%%%%%%%%%%%%%%%%%%%%%%%%%%%%%%%%%%%%%%%%%%%%%%%%
\subsection{Proof of Positive Root Condition: Simple $\mathfrak{g}$}
\label{simple}
When $\mathfrak{g}$ is simple, the range of the product  $(r,n';\vec{\ell}\,\,)>0$ in (\ref{Phidef}) is defined by  
\begin{align}
&n'r-\frac{1}{2}\,\vec{\ell}\cdot \vec{\ell}\geq -1\,,&&\text{and}
&&\left\{\begin{array}{ll} r>0,\, 
n'\in \mathbb{Z},\,\vec{\ell}\in\Lambda_{\mathfrak{g}} & \text{or} \\ 
r=0,\, n'> 0,\, \vec{\ell}\in\Lambda_{\mathfrak{g}}  & \text{or} \\ 
r=n'=0,\, \vec{\ell}\in\Lambda^+_{\mathfrak{g}} \
.\end{array}\right.\label{SumPosSimple}
\end{align} 
As mentioned above, the norm $||\cdot||^2$ in (\ref{LogTermInteg}) takes into account that there
are 
contributions with $(r,n';\vec{\ell}\,\,)>0$ and contributions with
$(r,n';\vec{\ell}\,\,)<0$. 
It remains to show that (\ref{SumPosSimple}) are the conditions that
characterise the elements of 
$\Lambda_{{\mathfrak g}^{++}}^+$ with $\alpha^2\leq 2$. Let us work with the 
set of simple roots $\alpha_I$ of $\mathfrak{g}^{++}$ in the following basis 
\begin{align}
&\alpha_{-1} = (1,-1;\vec{0} )\label{Singroot1} \\
&\alpha_{0} = (-1,0;-\vec{\theta}) \label{Singroot2}\\ 
&\alpha_{i} = (0, 0; \vec{e}_i) \ , \quad i=1,\ldots, k \ ,\label{Singroot3}
\end{align}
where $\vec{\theta}$ is the highest root of ${\mathfrak g}$ (which exists for every simple Lie algebra), and $\vec{e}_i$, $i=1,\ldots, k$ are the simple roots of ${\mathfrak g}$. We further parametrise the positive roots according to
An arbitrary positive root of $\mathfrak{g}^{++}$ may be written as a linear
combination of the simple roots  $\alpha_{-1}$, $\alpha_0, \alpha_i$
\begin{equation}
{\alpha}= 
\sum_{I=-1}^k
x_{I}\, {\alpha}_I\in\Lambda^+_{\mathfrak{g}^{++}}\ , \qquad \hbox{with} \qquad
 x_{I}\in\mathbb{Z}_+\ , \label{RootDecompgpp}
\end{equation}
where $\mathbb{Z}_+$ denotes the non-negative integers. Using the definition of
the inner product (\ref{innerProdRed}) we find that the scalar product of
$\alpha$ with $\y$ 
is given by
\begin{align}
({\alpha}|\y)=x_{-1}T+(x_0-x_{-1})U+(x_i\vec{e}_i-x_0\vec{\theta})\cdot\vec{V}\ .
\label{simpleinnerproduct}
\end{align}

Since the
exponent in 
(\ref{DenomFormPosRootwoWeyl})  is $c_{0}(-\alpha^2/2)$, it is natural to
identify 
\begin{equation}
x_{-1}=r\ , \qquad x_0=n'+r\ ,\qquad x_i\, \vec{e}_i=\vec{\ell}+(n'+r)\,
\vec{\theta} \ .
\end{equation}
Contracting the last identity with the fundamental weights $\vec{f}^i$ of
$\mathfrak{g}$ 
we can write the coefficients $x_i$ as $x_i=\vec{\ell}\cdot \vec{f}^i+(n'+r)\,
\vec{\theta}\cdot \vec{f}^i$. 
The proof then reduces to a case-by-case analysis. For example, if $r>0$, we
obviously  have
$x_{-1}>0$, but then in order for $n'r - \frac{1}{2} \vec{\ell} \cdot \vec{\ell}
\geq -1$, 
we need that $n'\geq -1$, thus  leading to $x_0\geq 0$. In order to understand
the condition
for $x_i$ we consider the different possibilities for $n'$ separately. If
$n'=-1$, then 
$r=1$ and $\vec{\ell}=\vec{0}$, and thus $x_i= 0$. Similarly,  for $n'=0$, 
$\vec{\ell} \cdot \vec{\ell} \leq 2$, which means that either $\vec{\ell}$ is a root of $\mathfrak{g}$ or $\vec{\ell}=\vec{0}$. In the latter case it follows
immediately that $x_i\geq 0$, 
while in the former case
\begin{align}
x_i= \vec{\ell}\cdot \vec{f}^i+r\, \vec{\theta}\cdot \vec{f}^i \geq
(r-1)\,\vec{\theta}\cdot \vec{f}^i\geq 0\,.
\end{align}
Finally, for $n'\geq 1$ we use the Cauchy-Schwarz inequality, following a
similar
discussion in \cite{Harvey:1996gc}, to conclude that 
\begin{align}
|\vec{\ell}\cdot  \vec{f}^i|^2\leq (\vec{f}^i\cdot \vec{f}^i)\, (\vec{\ell}\cdot
\vec{\ell}) 
\leq  (\vec{f}^i\cdot \vec{f}^i)\, (2+2n'r) \leq (n'+r)^2 \, (\vec{f}^i\cdot
\theta)^2 \ .
\end{align}
Since $\vec{\theta}\cdot \vec{f}^i\geq 0$ for all fundamental weights, it then
follows that also
$x_i\geq 0$. The other cases work similarly, and it follows that
(\ref{SumPosSimple}) 
characterises indeed the elements of $\Lambda_{{\mathfrak g}^{++}}^+$ with
$\alpha^2\leq 2$.
%%%%%%%%%%%%%%%%%%%%%%%%%%%%%%%%%%%%%%%%%%%%%%%%%%%%%%%%%%%%%%%%%%%%%%%%%%%%%%%%%
\subsection{Proof of Positive Root Condition: Semisimple $\mathfrak{g}$}
\label{semisimple}
Let us now repeat the discussion for the case that the broken gauge group $\mathfrak{g}$ is semisimple. For simplicity of presentation we shall restrict to the case when $\mathfrak{g}$ decomposes into a sum of two simple factors, $\mathfrak{g}=\mathfrak{g}_{(1)}\oplus\mathfrak{g}_{(2)}$, of rank $k_1$ and $k_2$, respectively. The generalisation to more factors is straight-forward. It is  still possible to write the integral in terms of an infinite product (\ref{LogTermInteg}),
 but now the condition on $(r,n';\vec{\ell})$ is replaced by the conditions 
\begin{align}
&n'r-\frac{1}{2}\,\vec{\ell}\cdot \vec{\ell}\geq -1
&&\text{and either}
&&\left\{\begin{array}{ll} r>0\,,\, n'\in
\mathbb{Z}\,,\,\vec{\ell}_{(1)}\in\Lambda_{\mathfrak{g}_{(1)}} \,\vec{\ell}_{
(2)}\in\Lambda_{\mathfrak{g}_{(2)}} & \\ r=0\,,\ n'>
0\,,\,\vec{\ell}_{(1)}\in\Lambda_{\mathfrak{g}_{(1)}}\,,\,\vec{\ell}_{(2)}
\in\Lambda_{\mathfrak{g}_{(2)}}\,, & \\ r=n'=0\,,\, \vec{\ell}\cdot
\Im\vec{V}>0\ ,\end{array}\right. \label{SumPosSemiSimple}
 \end{align}
where 
$\vec{\ell}\cdot \vec{\ell}=\vec{\ell}_{(1)}\cdot \vec{\ell}_{(1)}+\vec{\ell}_{(2)}\cdot \vec{\ell}_{(2)}$.  
Here we work in a chamber of the moduli space where 
\begin{align}
\Im\vec{V}\in \left(\Lambda^+_{\mathfrak{g}_{(1)}}\oplus
\Lambda^+_{\mathfrak{g}_{(2)}}\right) \otimes \mathbb{C}\ ,
\label{CondModSpaceSemi}
\end{align} 
such that the only contribution to the degenerate orbit with $\vec{\ell}\neq \vec{0}$ 
comes from vectors $\vec{\ell}$ which correspond to simple roots of either 
$\mathfrak{g}_{(1)}$ or $\mathfrak{g}_{(2)}$, both of which have  length squared
two. %$\vec{e}_i\cdot \vec{e}_i=2$.  
We will now show that (\ref{SumPosSemiSimple}) are just the conditions which characterise
`positive' elements of the root lattice of $\mathfrak{g}^{++}$ of norm $\alpha^2\leq 2$, which
--- just as in the simple case --- will allow us to reinterpret
$\Phi_{\mathfrak{g}}$ as an infinite product of the form (\ref{DenomFormPosRootwoWeyl})
over the positive roots of $\bor((\mathfrak{g}_{(1)}\oplus\mathfrak{g}_{(2)})^{++})$.\footnote{See Appendix A of \cite{GHPI} for our conventions for the double extension $(\mathfrak{g}_{(1)}\oplus\mathfrak{g}_{(2)})^{++}$.}
%$\Delta^+_{\bor(\mathfrak{g}_{(1)}\oplus\mathfrak{g}_{(2)})}$.
%\begin{align}
%\Phi_{\mathfrak{g}}(y)=\prod_{\alpha
%\in\Delta^+_{\bor(\mathfrak{g}_{(1)}\oplus\mathfrak{g}_{(2)})}}\,
% \left(1-e^{2\pi i (\alpha|y)}\right)^{c_{[0]}(-\alpha^2/2)}\
%.\label{DenomFormPosRootwoWeylSemi}
%\end{align}
Here the `positive'  elements of the root lattice of $\mathfrak{g}^{++}$ are those
that have positive scalar product with a fixed vector $\beta$ of the underlying vector space 
\begin{align}
\Lambda^+_{\mathfrak{g}^{++}}=\left\{x\in\Lambda_{\mathfrak{g}^{++}}
:(x|\beta)>0\right\}\ .
\end{align}
For further convenience we will choose the vector $\vec{\beta}$ to be of the
form 
\begin{align}
&\beta=\left(u+2,u+1;\we;\wz\right) &&\text{with}&&u=\vec{\theta}_1\cdot
\we+\vec{\theta}_2\cdot \wz>0\ ,
\end{align}
where $\wv=(\we,\wz)\in \Lambda_{\mathfrak{g}_{(1)}}^+\oplus \Lambda_{\mathfrak{g}_{(2)}}^+$. 
In the following we will find it useful to introduce
$\vec{\theta}=(\vec{\theta}_{(1)},\vec{\theta}_{(2)})\in\Lambda_{\mathfrak{g}}$. Let us also introduce 
a basis of simple roots $\tilde{\alpha}_I$ for
${\tilde{\mathfrak{g}}^{++}}$ 
\begin{align}
&{\tilde{\alpha}}_{-1}=(1,-1;\vec{0};\vec{0})
&&{\tilde{\alpha}}_0^{(1)}=(-1,0;-\vec{\theta}_{(1)};\vec{0})
&&{\tilde{\alpha}}_0^{(2)}=(-1,0;\vec{0};-\vec{\theta}_{(2)})\label{SetRootsOver1}\\
%\label{SetRootsOver3}\\
&{\tilde{\alpha}}_i^{(1)}=(0,0;\vec{e}^{(1)}_i;\vec{0}) 
&&{\tilde{\alpha}}_m^{(2)}=(0,0;\vec{0};\vec{e}^{(2)}_m) \ ,
&& \label{SetRootsOver5}
\end{align}
where $i=1,\ldots, k_1$, $m=1,\ldots,k_2$, and $\vec{e}^{\,(1)}, \vec{e}^{\,(2)}$ 
are simple roots of $\mathfrak{g}_{(1)}, \mathfrak{g}_{(2)}$, with 
$\vec{\theta}_{(1)}, \vec{\theta}_{(2)}$ the corresponding highest roots.
The roots (\ref{SetRootsOver1}), (\ref{SetRootsOver5}) define an overcomplete basis for the root lattice
$\Lambda_{{\mathfrak{g}^{++}}}=\Pi^{1,1}\oplus\Lambda_{\mathfrak{g}_{(1)}}
\oplus\Lambda_{\mathfrak{g}_{(2)}}$. 
%An arbitrary root can then be written as
%\begin{align}
%&\alpha=\tilde{x}_{-1}{\tilde{\alpha}}_{-1}+\tilde{x}^{(1)}_0\tilde{\alpha}_0^{
%(1)}+\tilde{x}^{(2)}_0\tilde{\alpha}_0^{(2)}+\sum_{i=1}^{k_1}\tilde{x}_i^{(1)}{
%\tilde{\alpha}}_i^{(1)}+\sum_{m=1}^{k_2}\tilde{x}_m^{(2)}{\tilde{\alpha}}_m^{(2)
%}\,.\label{SemiSimplePosRoot}
%\end{align}
In fact, there is one relation (generating the center $\mathfrak{r}$ of ${\tilde{\mathfrak{g}}^{++}}$, 
see \cite{GHPI} for more details) which we may use to express 
$\tilde{\alpha}^{(2)}_0$ in terms of the other roots 
\begin{align}
\tilde{\alpha}^{(2)}_0=\tilde{\alpha}^{(1)}_0
+\sum_{i=1}^{k_1}(\vec{\theta}_{(1) }\cdot \vec{f}^i)\, \tilde{\alpha}^{(1)}_i
-\sum_{m=1}^{k_2}(\vec{\theta}_{(2)}\cdot \vec{f}^m)\, \tilde{\alpha}^{(2)}_m\ .
\label{SemiSimpExpressRoot}
\end{align}
Here $\vec{f}^i$ and $\vec{f}^m$ are the fundamental weights of
$\mathfrak{g}_{(1)}$ and $\mathfrak{g}_{(2)}$, respectively. With this relation
we can then write for any $\alpha\in\Lambda_{\mathfrak{g}^{++}}$
\be
\alpha=x_{-1}{\tilde{\alpha}}_{-1}+x_0\tilde{\alpha}_0^{(1)}+\sum_{i=1}^{k_1}x_i^{
(1)}{\tilde{\alpha}}_i^{(1)}+\sum_{m=1}^{k_2}x_m^{(2)}{\tilde{\alpha}}_m^{(2)}.
\label{GenElRootSem}
\ee
Using the same inner product as in (\ref{innerProdRed}) we find that the product
between $\alpha$ and a moduli vector $\y=(U,T;\vec{V}_{(1)}, \vec{V}_{(2)})$
reads
\begin{align}
({\alpha}|\y)=x_{-1}T+(x_0-x_{-1})U&+\left(\sum_{i=1}^{k_1}x_i^{(1)}\vec{e}
_i-x_0\,\vec{\theta}_{(1)}\right)\cdot\vec{V}_{(1)}+\sum_{m=1}^{k_2}x_m^{(2)}
\vec{e}_m\cdot\vec{V}_{(2)}\,.\label{semisimpleinnerproduct}
\end{align}

With these preparations the scalar product of a generic vector
$\alpha\in\Lambda_{\mathfrak{g}^{++}}$, parametrised as in (\ref{GenElRootSem}),
with $\beta$ is given by
\begin{align}
(\alpha|\beta)=x_{-1}+x^{0}(u+1)+\left(\sum_{i=1}^{k_1}x_i^{(1)}\vec{e}_i-x_0\,
\vec{\theta}_{(1)}\right)\cdot\we+\sum_{m=1}^{k_2}x_m^{(2)}\vec{e}_m\cdot\wz\ .
\label{GenFormProjPosVecCon}
\end{align}
Comparing (\ref{semisimpleinnerproduct}) to the exponent of the
denominator formula (\ref{DenomFormPosRootwoWeyl}) suggests the identification
\begin{align}
&x_{-1}=r &&x_0=n'+r &&\sum_{i=1}^{k_1}x_i^{(1)}\vec{e}_i-\vec{\theta}_{(1)}
(n'+r)=\vec{\ell}_{(1)} &&\sum_{m=1}^{k_2}x_m^{(2)}\vec{e}_m=\vec{\ell}_{(2)}\ ,\nonumber
\end{align}
in terms of which the scalar product (\ref{GenFormProjPosVecCon}) becomes
\begin{align}
(\alpha|\beta)=r+(n'+r)(u+1)+(\vec{\ell}\cdot\wv)\,.\label{StreamProj}
\end{align}
In order to show that (\ref{SumPosSemiSimple}) indeed characterises vectors of
$\Lambda^+_{\mathfrak{g}^{++}}$ with norm $\leq 2$, we first have to show that 
(\ref{StreamProj}) is positive for all three cases in (\ref{SumPosSemiSimple}).
This can again be done by a case-by-case analysis which is rather 
similar to that in section~\ref{Sect:DenominatorFormula}. For example, for $r>0$ 
the first equation implies $n'\geq -1$. If $n'=-1$ the first equation furthermore implies that 
$\vec{\ell}=\vec{0}$, and thust $(\alpha|\beta)>0$. For $n'=0$ we have instead
$\vec{\ell}\cdot\vec{\ell}\leq 2$ which means that either $\vec{\ell}=\vec{0}$ or 
$\vec{\ell}$ is one of the roots of $\mathfrak{g}_{(1)}$ or $\mathfrak{g}_{(2)}$. 
In the former case it immediately follows that $(\alpha|\beta)=r(u+2)>0$, while in the latter case 
\begin{align}
(\alpha|\beta)\geq
r(u+2)-\text{max}\left(\vec{\theta}_{(1)}\cdot\we,\vec{\theta}_{(2)}
\cdot\wz\right)>0\ .
\end{align}
Finally, for $n'>0$ we can estimate
\begin{align}
(\alpha|\beta)&\geq
2r+n'+(n'+r)(\vec{\theta}\cdot \wv)-|\vec{\ell}\cdot\wv|\geq 2r+n'+(n'+r)(\vec{\theta}\cdot \wv)
-\sqrt{(\wv\cdot\wv)(\vec{\ell}\cdot\vec{\ell})}\nonumber\\
&\geq 2r+n'+(n'+r)(\vec{\theta}\cdot \wv)
-\sqrt{(2+2n'r)(\wv\cdot\wv)}>0\,,\label{SemiCauchyUse}
\end{align}
where the last inequality follows from expanding $\wv$ into weights $\vec{f}_i$
for $\mathfrak{g}_{(1)}$ and $\mathfrak{g}_{(2)}$, respectively, and using the
estimate $(\vec{\theta}\cdot \vec{f}_i)^2(\vec{\theta}\cdot \vec{f}_j)^2\geq
(\vec{f}_i\cdot\vec{f}_i)(\vec{f}_j\cdot\vec{f}_j)\geq (\vec{f}_i\cdot
\vec{f}_j)^2$. All other cases follow in a similar fashion and we will not explicitly write them down here. 

Conversely, one can also show that if (\ref{SumPosSemiSimple}) is not satisfied, 
the corresponding $\alpha$ is not an element of $\Lambda_{\mathfrak{g}^{++}}^+$ 
since $(\alpha|\beta)<0$. Thus, also in the case 
$\mathfrak{g}$ being semisimple, (\ref{LogTermInteg}) can be written in the form of
(\ref{DenomFormPosRootwoWeyl}) which is identified with the infinite
product part of the denominator formula for the Borcherds algebra
$\bor(\mathfrak{g}^{++})$. 
\section{Jacobi Forms for $\Gamma_0(4)$}\label{App:ModFormsG04}
For the explicit computations in section~\ref{Sect:ExA1} we require some terminology of weak Jacobi forms (in the framework of the congruence subgroup $\Gamma_0(4)$). Any weak Jacobi form of index $1$ can be expanded in terms of a basis of Jacobi forms of weight $0$ and $-2$ respectively \cite{EZ} 
\begin{align}
f_{w,1}(\tau,z)=h^{(1)}_w(\tau)\,\phi_{0,1}(\tau,z)+h^{(2)}_{w+2}(\tau)\,\phi_{-2,1}(\tau,z)\,,
\end{align}
where we have the definitions
\begin{align}
&\phi_{0,1}(\tau,z):=4\sum_{i=2}^4\frac{\vartheta_i(\tau,z)^2}{\vartheta_i(\tau,0)^2}\,,&&\phi_{-2,1}(\tau,z):=-\frac{\vartheta_1(\tau,z)^2}{\eta(\tau)^6}\,,
\end{align}
and $h^{(1,2)}$ are modular forms of weight $w$ and $w+2$ respectively. In section~\ref{Sect:ExA1} we will be interested in the case where the latter are not modular forms under the full $SL(2,\mathbb{Z})$, but rather one of its congruence subgroups $\Gamma_0(N)$, where we define for $N\in \mathbb{N}$
\begin{align}
\Gamma_0(N):=\left\{\left(\begin{array}{cc}a & b \\ c & d\end{array}\right)\in SL(2,\mathbb{Z}):c=0\hspace{0.2cm}\text{mod}\hspace{0.2cm}N\right\}\,.
\end{align}
Specifically, we will be interested in the case $w=4$ and $N=4$. A (for our purposes convenient) basis for the spaces $M_w(\Gamma_0(4))$ of modular forms of weight $w$ under $\Gamma_0(4)$ is given by (for further details see~\cite{Gaberdiel:2010ca})
\begin{align}
&M_4(\Gamma_{0}(4)):\,\{E_4(\tau),E_4(2\tau),E_4(4\tau)\}\,,\\
&M_6(\Gamma_{0}(4)):\,\{E_6(\tau),E_6(2\tau),E_6(4\tau),h_6(\tau)\}\,,
\end{align}
where $h_6(\tau)$ is an element of the space of cusp forms (\emph{i.e.} forms which vanish at all cusps of $\overline{\mathbb{H}/\Gamma_0(N)}$). Its Fourier expansion is given by\footnote{This can be extracted from {\tt http://modi.countnumber.de/}.}
\begin{align}
h_6(\tau)=q - 12q^3 + 54q^5 - 88q^7 - 99q^9 + 540q^{11}-418q^{13} - 648q^{15} + 594q^{17} + \mathcal{O}(q^{19})\,.
\end{align}

%%%%%%%%%%%%%%%%%%%%%%%%%%%%%%%%%%%%%%%%%%%%%%%%%%%%%%%%%%%%%%%%%%%%%%%%%%%%%%%%%%%%%%%%%%%%%%%%%%%%%%%%


\begin{thebibliography}{99}


\bibitem{GHPI}
  M.~R.~Gaberdiel, S.~Hohenegger, D.~Persson,
  \emph{Borcherds Algebras and N=4 Topological Amplitudes},
  JHEP {\bf 1106 } (2011)  125.
 {\tt  [arXiv:1102.1821 [hep-th]]}.
  
\bibitem{Antoniadis:2006mr} 
I.~Antoniadis, S.~Hohenegger and K.S.~Narain, 
{\it N = 4 topological amplitudes and string effective action}, 
Nucl.\ Phys.\  B {\bf 771} (2007) 40 
{\tt [arXiv:hep-th/0610258]}.
%%CITATION = NUPHA,B771,40;%%

\bibitem{Harvey:1995fq} 
J.A.~Harvey and G.W.~Moore, 
{\it Algebras, BPS states, and strings}, 
Nucl.\ Phys.\  B {\bf 463} (1996) 315 
\texttt{[arXiv:hep-th/9510182]}.
%%CITATION = NUPHA,B463,315;%%

\bibitem{Harvey:1996gc}
J.A.~Harvey and G.W.~Moore,
{\it On the algebras of BPS states},
Commun.\ Math.\ Phys.\  {\bf 197} (1998) 489
\texttt{[arXiv:hep-th/9609017]}.
%%CITATION = CMPHA,197,489;%%

\bibitem{Borcherds1}
R.E.~Borcherds, 
{\it Automorphic forms on $O_{s+2,2}(\mathbb{R})$ and infinite products}, 
Invent.\ Math.\ \textbf{120} (1995) 161.

\bibitem{Borcherds2}
R.E.~Borcherds, 
{\it Automorphic forms with singularities on Grassmannians},
Invent.\ Math.\ {\bf 132} (1998) 491.


\bibitem{Antoniadis:2007cw} 
I.~Antoniadis, S.~Hohenegger, K.S.~Narain and E.~Sokatchev,
{\it Harmonicity in N=4 supersymmetry and its quantum anomaly}, 
Nucl.\ Phys.\  B {\bf 794} (2008) 348 
{\tt [arXiv:0708.0482 [hep-th]]}.
%%CITATION = NUPHA,B794,348;%%

\bibitem{Berkovits:1994vy}
N.~Berkovits and C.~Vafa,
{\it N=4 topological strings},
Nucl.\ Phys.\  B {\bf 433} (1995) 123 
\texttt{[arXiv:hep-th/9407190]}.
 %%CITATION = NUPHA,B433,123;%%

\bibitem{Ooguri:1991fp} 
H.~Ooguri and C.~Vafa, 
{\it Geometry of N=2 strings},
Nucl.\ Phys.\  B {\bf 361} (1991) 469.
%%CITATION = NUPHA,B361,469;%%

\bibitem{Dixon:1990pc} 
L.J.~Dixon, V.~Kaplunovsky and J.~Louis, 
{\it Moduli dependence of string loop corrections to gauge coupling constants}, 
Nucl.\ Phys.\  B {\bf 355} (1991) 649.
%%CITATION = NUPHA,B355,649;%%

\bibitem{Henningson:1996jz}
M.~Henningson and G.W.~Moore, {\it Threshold corrections in K3 x T2 
heterotic string compactifications,} Nucl.\ Phys.\  B {\bf 482} (1996) 187 
\texttt{[arXiv:hep-th/9608145]}.

\bibitem{Marino:1998pg} 
M.~Marino and G.W.~Moore, 
{\it Counting higher genus curves in a Calabi-Yau manifold}, 
Nucl.\ Phys.\  B {\bf 543} (1999) 592 
{\tt [arXiv:hep-th/9808131]}.
%%CITATION = NUPHA,B543,592;%%

\bibitem{Weiss:2007tk} 
M.~Weiss, 
{\it Topological amplitudes in heterotic strings with Wilson lines}, 
JHEP {\bf 0708} (2007) 024 
{\tt [arXiv:0705.3112 [hep-th]]}.
%%CITATION = JHEPA,0708,024;%%

\bibitem{LopesCardoso:1996nc} 
G.~Lopes Cardoso, G.~Curio and D.~Lust, 
{\it Perturbative couplings and modular forms in N = 2 string models with a
Wilson line}, 
Nucl.\ Phys.\  B {\bf 491} (1997) 147 
{\tt [arXiv:hep-th/9608154]}.
%%CITATION = NUPHA,B491,147;%%

\bibitem{Stieberger:1998yi} 
S.~Stieberger, 
{\it (0,2) heterotic gauge couplings and 
their M theory origin,} Nucl.\ Phys.\  B {\bf 541} (1999)  109
\texttt{[arXiv:hep-th/9807124]}.

\bibitem{Cheng:2008fc}
M.C.N.~Cheng and E.P.~Verlinde,
{\it Wall crossing, discrete attractor flow, and Borcherds algebra}, SIGMA {\bf 4} (2008) 068 \texttt{[arXiv:0806.2337 [hep-th]]}.
%CITATION = 00480,4,068;%%

\bibitem{Gritsenko:1996ax} 
V.A.~Gritsenko and V.V.~Nikulin, 
{\it The Igusa modular forms and `the simplest' Lorentzian Kac--Moody algebras},
{[\tt arXiv:alg-geom/9603010]}.
%%CITATION = ALG-GEOM/9603010;%%

\bibitem{GritsenkoNikulin}
V.A.~Gritsenko and V.V.~Nikulin, 
{\it Siegel automorphic form corrections to some Lorentzian Kac-Moody algebras},
C.\ R.\ Acad.\ Sci.\ Paris S\'er.\ A--B.\ {\bf 321} (1995) 1151.

\bibitem{Lerche:1999ju}
W.~Lerche and S.~Stieberger,
{\it 1/4 BPS states and non-perturbative couplings in N = 4 string theories},
Adv.\ Theor.\ Math.\ Phys.\  {\bf 3} (1999) 1539
{\tt  [arXiv:hep-th/9907133]}.
%%CITATION = 00203,3,1539;%%
  
%\cite{Lust:1989tj}
\bibitem{Lust:1989tj}
  D.~Lust and S.~Theisen,
  ``Lectures on string theory,''
  Lect.\ Notes Phys.\  {\bf 346} (1989) 1.
  %%CITATION = LNPHA,346,1;%%
  
\bibitem{Gannon:1991vg} 
T.~Gannon and C.S.~Lam, 
{\it Lattices and theta function identities. 1. Theta constants}, 
J.\ Math.\ Phys.\  {\bf 33} (1992) 854.
%%CITATION = JMAPA,33,854;%%

\bibitem{Antoniadis:2007ta} I.~Antoniadis and S.~Hohenegger, 
{\it Topological amplitudes and physical couplings in string theory,}
Nucl.\ Phys.\ Proc.\ Suppl.\  {\bf 171 } (2007)  176 
{\tt [arXiv:hep-th/0701290]}.

\bibitem{Antoniadis:1995zn} 
I.~Antoniadis, E.~Gava, K.S.~Narain and T.R.~Taylor, 
{\it N=2 type II heterotic duality and higher derivative F terms}, 
Nucl.\ Phys.\  B {\bf 455} (1995) 109 
{\tt [arXiv:hep-th/9507115]}.
%%CITATION = NUPHA,B455,109;%%

\bibitem{Antoniadis:2010iq} 
I.~Antoniadis, S.~Hohenegger, K.S.~Narain and T.R.~Taylor, 
{\it Deformed topological partition function and Nekrasov backgrounds}, 
Nucl.\ Phys.\ B {\bf 838} (2010) 253
{\tt [arXiv:1003.2832 [hep-th]]}.
%%CITATION = ARXIV:1003.2832;%%

\bibitem{Dijkgraaf}
R.~Dijkgraaf, {\it Mirror symmetry and elliptic curves},
in: ``The moduli space of curves'', Prog.\ Math.\ {\bf 129} (1995) 149.

\bibitem{KanekoZagier}
M.~Kaneko and D.~Zagier, 
{\it A generalized Jacobi theta function and quasi-modular forms},
in: ``The moduli space of curves'', Prog.\ Math.\ {\bf 129} (1995) 165.

%\cite{Kiritsis:1997hf}
\bibitem{Kiritsis:1997hf}
  E.~Kiritsis and N.~A.~Obers,
  ``Heterotic type I duality in D < 10-dimensions, threshold corrections and D
  instantons,''
  JHEP {\bf 9710}, 004 (1997)
 {\tt  [arXiv:hep-th/9709058]}.
  %%CITATION = JHEPA,9710,004;%%

\bibitem{Lerche:1998nx} 
W.~Lerche and S.~Stieberger, 
{\it Prepotential, mirror map and F theory on K3,} 
Adv.\ Theor.\ Math.\ Phys.\  {\bf 2} (1998)  1105
\texttt{[arXiv:hep-th/9804176]}.

\bibitem{Foerger:1998kw} 
K.~Foerger and S.~Stieberger, 
{\it Higher derivative couplings and heterotic type I duality in eight-dimensions,} 
Nucl.\ Phys.\  B {\bf 559} (1999)  277 
\texttt{[arXiv:hep-th/9901020]}.


\bibitem{Obers:1999um}
N.A.~Obers and B.~Pioline,
{\it Eisenstein series and string thresholds},
Commun.\ Math.\ Phys.\  {\bf 209} (2000) 275
{\tt [arXiv:hep-th/9903113]}.
%%CITATION = CMPHA,209,275;%%
 
 \bibitem{EZ} 
M.~Eichler and D.~Zagier, 
{\it The Theory of Jacobi Forms}, 
Birkh\"auser (1985).

\bibitem{Kontsevich}
M.~Kontsevich, {\it Product formulas for modular forms on $O(2,n)$ (after R. Borcherds)}, 
S\'eminaire Bourbaki, Vol. 1996/1997, Ast\'erisque No. 245 (1997), \texttt{[arXiv:alg-geom/9709006]}

\bibitem{Prasad}
D.~Prasad, \emph{A brief survey on the theta correspondence},
Lectures given at Trichy in January 1996, available from
{\texttt{http://www.math.tifr.res.in/$\,  \widetilde{}\, $dprasad/dp.pdf}}

\bibitem{Howe}
R.~Howe, \emph{$\theta$-series and invariant theory},
Proc. Symp. Pure Math. {\bf 33}, ÊPart 1, 275 (1979).
 
\bibitem{Gunning} C.~Gunning, {\it Lectures on Modular Forms,} Princeton University Press, (1962). 
 
\bibitem{Antoniadis:2009tr} 
I.~Antoniadis and S.~Hohenegger, 
{\it N=4 Topological amplitudes and black hole entropy}, 
Nucl.\ Phys.\ B {\bf 837} (2010) 61
{\tt [arXiv:0910.5596 [hep-th]]}.
%%CITATION = ARXIV:0910.5596;%% 
 
 \bibitem{FunkeBruinier}
J.~H.~Bruinier and J.~Funke, {\it On the injectivity of the Kudla-Millson lift and surjectivity of the 
Borcherds lift}, \texttt{[arXiv:math/0606178]}.

 
%\cite{Moore:1997pc}
\bibitem{Moore:1997pc}
  G.~W.~Moore and E.~Witten,
  \emph{Integration over the u plane in Donaldson theory},
  Adv.\ Theor.\ Math.\ Phys.\  {\bf 1 } (1998)  298.
 \texttt{[arXiv:hep-th/9709193]}.
 
 %\cite{Kiritsis:2000zi}
\bibitem{Kiritsis:2000zi}
  E.~Kiritsis, N.~A.~Obers and  B.~Pioline,
  \emph{Heterotic / type II triality and instantons on K(3)},
  JHEP {\bf 0001 } (2000)  029.
 {\tt [hep-th/0001083]}.

\bibitem{Bruinier}
J.~H.~Bruinier, \emph{Hilbert modular forms and their applications},
\texttt{arXiv:math/0609763v1} 

\bibitem{GritsenkoClery}
V. Gritsenko and F. Clery, \emph{he Siegel modular forms of genus 2 with the simplest divisor},
{\tt [arXiv:0812.3962 [math.NT]]}.

\bibitem{Borcherds0}
R.E.~Borcherds, 
{\it Generalized Kac-Moody algebras}, 
J.\ Algebra \textbf{115} (1988) 501.

\bibitem{Aldazabal:1995yw} 
G.~Aldazabal, A.~Font, L.E.~Ibanez and F.~Quevedo, 
{\it Chains of N=2, D=4 heterotic/type II duals},
Nucl.\ Phys.\  B {\bf 461} (1996) 85
\texttt{[arXiv:hep-th/9510093]}.
%%CITATION = NUPHA,B461,85;%%

\bibitem{Gannon:1991vj} 
T.~Gannon and C.S.~Lam, 
{\it Lattices and theta function identities. 2. Theta series}, 
J.\ Math.\ Phys.\  {\bf 33} (1992) 871.
%%CITATION = JMAPA,33,871;%%

\bibitem{Conway} 
J.H.~Conway and N.J.A.~Sloane, 
{\it Sphere Packings, Lattices and Groups}, 
Springer (1998).

%\cite{Kleinschmidt:2008jj}
\bibitem{Kleinschmidt:2008jj}
  A.~Kleinschmidt and  D.~Roest,
  \emph{Extended Symmetries in Supergravity: The Semi-simple Case},
  JHEP {\bf 0807 } (2008)  035.
{\tt  [arXiv:0805.2573 [hep-th]]}.

%\cite{Michel:2008bx}
\bibitem{Michel:2008bx} Y.~Michel, B.~Pioline and C.~Rousset, {\it N=4 BPS black holes and octonionic twistors,} JHEP {\bf 0811} (2008) 068 {\tt [arXiv:0806.4563 [hep-th]]}.
  %%CITATION = JHEPA,0811,068;%%

\bibitem{Govindarajan:2008vi}
S.~Govindarajan and K.~Gopala Krishna,
{\it Generalized Kac-Moody algebras from CHL dyons},
JHEP {\bf 0904} (2009) 032 
\texttt{[arXiv:0807.4451 [hep-th]]}.
%%CITATION = JHEPA,0904,032;%%

S.~Govindarajan and K.~Gopala Krishna,
{\it BKM Lie superalgebras from dyon spectra in $\mathbb{Z}_N$ CHL orbifolds for
composite N},
JHEP {\bf 1005} (2010) 014 
\texttt{[arXiv:0907.1410 [hep-th]]}.
 %%CITATION = JHEPA,1005,014;%%


S.~Govindarajan, 
{\it BKM Lie superalgebras from counting twisted CHL dyons}, 
\texttt{[arXiv: 1006.3472 [hep-th]]}.
%%CITATION = ARXIV:1006.3472;%%

\bibitem{Gaberdiel:2010ca} M.~R.~Gaberdiel, S.~Hohenegger and R.~Volpato, {\it Mathieu Moonshine in the elliptic genus of K3,} JHEP {\bf 1010 } (2010)  062. {\tt [arXiv:1008.3778 [hep-th]]}.





\end{thebibliography}
\end{document}